\documentclass[twocolumn]{aastex631}
\usepackage{amsmath}

\begin{document}

\title{Cosmic Ray Magnetohydrodynamics: A New Two-Moment Framework with Numerical Implementation}

\newcommand{\xzhao}[1]{{\color[rgb]{1, 0, 0}{#1}}}

\author[0009-0003-1259-319X]{Xihui Zhao}
\affiliation{Institute for Advanced Study, Tsinghua University, Beijing 100084, China \\}
\affiliation{Department of Astrophysical Sciences, Princeton University, Princeton, NJ 08544, USA \\
}

\author[0000-0001-6906-9549]{Xue-Ning Bai}
\affiliation{Institute for Advanced Study, Tsinghua University, Beijing 100084, China \\}
\affiliation{Department of Astronomy, Tsinghua University, Beijing 100084, China \\
}

\author[0000-0002-0509-9113]{Eve C. Ostriker}
\affiliation{Department of Astrophysical Sciences, Princeton University, Princeton, NJ 08544, USA \\
}



\begin{abstract}
Cosmic rays (CRs) play a pivotal role in various astrophysical systems, delivering feedback over a broad range of scales. However, modeling CR transport remains challenging due to its inherently multi-scale nature and complex microphysics. Recent advances in two-moment CR hydrodynamics have alleviated some of these challenges, improving our understanding of CR feedback. Yet, current two-moment methods may not be able to directly incorporate all relevant CR transport processes, while the outcome of CR feedback sensitively depends on these underlying microphysics. Furthermore, numerical challenges persist, including instabilities from streaming terms and ambiguities in solver design for coupled CR-MHD systems. In this work, we develop a two-moment description for CR hydrodynamics from first principles. Beyond canonical CR streaming, our formulation accounts for CR pressure anisotropy and Alfv\'en waves propagating in both directions along the magnetic field, providing a general framework to incorporate more CR transport physics. We implement this framework as a new CR fluid module in the \textit{Athena}++ code, and validate it through a suite of benchmark tests. In particular, we derive the full dispersion relation of the two-moment CR-MHD system, identifying the CR-acoustic instability as well as other wave branches. These CR-MHD waves serve as rigorous benchmarks and also enable the use of realistic signal speeds in our Riemann solver. We propose a time step guideline to mitigate numerical instabilities arising from streaming source terms.

\end{abstract}



\section{Introduction} 

Cosmic rays (CRs) are high-energy charged particles traversing the Universe at velocities close to the speed of light. Despite their relativistic speeds, CRs do not freely escape from the galactic disk at the light-crossing time scale, but are instead confined for a time scale $\sim 10^4$ times longer, leading to CR energy densities roughly in equipartition with other components of the interstellar medium (ISM), including thermal, turbulent, radiation and magnetic energy densities in galaxies \citep{Boulares1990,Ferriere2001,Elmegreen_Scalo2004,Cox2005,grenier2015}. Effects of CRs are thus considered as an integral part of ISM dynamics \citep[see][for reviews]{Zweibel17,Ruszkowski23,Hopkins2025}. Recent advances in observational and computational techniques have renewed interest in CRs due to their known or potential roles in ionizing gas \citep{Padovani2009}, regulating star formation \citep[e.g.,][]{Jubelgas2008,Chen2016,Farcy2022}, driving galactic winds \citep[for very recent work, see e.g.,][additional earlier studies are listed below]{Sike_2025,Thomas_2025}, shaping the phase structure in the circumgalactic medium (CGM) \citep[e.g.,][]{Sharma2010,Ji2020,HuangXS2022,Weber2025}, and modulating black hole feedback \citep[e.g.,][]{Sijacki2008,Guo2008,Lin2023,Su2025}. It has become essential for studies of galaxy formation and evolution to  incorporate CR feedback properly, as CRs are actively involved in a vast array of processes in the life cycle of a galactic ecosystem.

Modeling CR feedback numerically, however, remains challenging due to its multi-scale nature. For $\sim \rm GeV$ CRs, which dominate the CR energy distribution and are dynamically the most important, gyroradii are typically $\lesssim 1$AU (or $\sim 10^{-6}$pc). Directly modeling both microphysical kinetics and galactic scales is prohibitive, given the dynamic range spanning 10 orders of magnitude \citep{Zweibel17}. To study large scale one therefore must coarse-grain the kinetic physics of CRs, and treat CR ensembles as a bulk fluid. The fluid formulation for CRs was initially developed by integrating over momentum the Fokker-Plank equation that describes CR transport \citep[e.g., as in][]{Skilling1971,Skilling1975,Schlickeiser02,Zank2014}. Assuming a nearly isotropic CR distribution, an energy-weighted integration leads to a fluid equation governing the evolution of CR energy density \citep[e.g., as in][]{Drury1981,McKenzie1982,Voelk1984,Breit1991,Pfrommer2017}:

\begin{equation}\label{eq:one_moment}
    \begin{aligned}
        \frac{\partial \mathcal{E}_{\rm cr}}{\partial t}&+\nabla \cdot[(\vec{u}+\vec{v}_{\rm st})(\mathcal{E}_{\rm cr}+P_{\rm cr})]=&\\&(\vec{u}+\vec{v}_{\rm st})\cdot \nabla P_{\rm cr}+\nabla\cdot[\kappa\vec{b}(\vec{b}\cdot\nabla P_{\rm cr})]&
    \end{aligned}
\end{equation}
Here $\mathcal{E}_{\rm cr}$, $P_{\rm cr}\equiv \mathcal{E}_{\rm cr}/3$ and $\kappa$ represent CR energy density, CR pressure, and the CR diffusion coefficient, respectively. $\vec{u}$ denotes the MHD gas velocity and $\vec{b}\equiv \vec{B}/|\vec{B} |$ is the unit vector along the large-scale background magnetic field. $\vec{v}_A \equiv \vec{B}/(4 \pi \rho)^{1/2}$ is the traditional Alfv\'en velocity and $\vec{v}_{\rm st}\equiv -{\rm sgn}(\vec{b}\cdot\nabla P_{\rm cr})\vec{v}_A$ is the streaming velocity, which points down the gradients in the CR pressure. 

Coupling the one-moment CR hydrodynamic equation, \autoref{eq:one_moment}, with MHD equations enabled early CR feedback models that demonstrated the possibility of CRs driving galactic winds \citep[e.g.,][]{Ipavich1975,Breit1991,Zirakashvili1996,Samui2010,Dorfi2012,Recchia2017,Mao2018,Quataer2022a,Quataer2022b}. However, this one-moment equation is known to induce numerical instabilities through the streaming velocity term $\vec{v}_{\rm st}$, which is ill-defined and flips sign  at $P_{\rm cr}$ extrema. To address this step-function-like discontinuity, \cite{Sharma10} added a numerical diffusion term to regularize the sharp transition, but at the cost of involving more computational expense and artificial diffusion. 

Drawing parallels with radiation hydrodynamics (RHD), \cite{JiangOh2018} developed an alternative two-moment framework that additionally evolves a separate equation for CR energy flux (see also \cite{Chan2019} for a similar implementation). Independently, \cite{Thomas19} derived a similar two-moment formalism from first principles based on a moment hierarchy of the CR Fokker-Planck equation. They additionally evolve a wave subsystem to account for the coupling between CRs and the MHD gas. By incorporating a reduced speed of light to accelerate integration \citep[as developed for the RHD context by, e.g.][]{Gnedin_Abel2001,Skinner2013}, two-moment methods retain computational tractability and have become the new norm in CR feedback modeling.
Despite their significant advantages over previous numerical approaches, two-moment methods for CR hydrodynamics still have limitations. Perhaps chief among these are the uncertainties in the microphysical CR transport model, and the challenges of integrating kinetic effects into fluid-scale descriptions to accurately reflect macroscopic energy and momentum transfer.

Due to their low number densities and relativistic speeds, GeV CRs are nearly collisionless particles, minimally interacting with the background ions, atoms, and electrons. Instead, they exchange energy and momentum efficiently with the thermal plasmas through resonant scattering by small-scale electromagnetic fluctuations, whose wavelengths are comparable to the CR gyro-radii \citep{Ruszkowski23}. Origins of such magnetic fluctuations include cascade of large-scale MHD turbulence and MHD waves driven by plasma instabilities. While high-energy CRs $(\gtrsim 100{\rm GeV})$ are largely confined by external turbulence \citep[e.g.,][]{Chandran2000,Yan2011}, lower-energy CRs—most relevant to ISM dynamics and chemistry—are primarily scattered by MHD waves generated by CRs themselves via gyro-resonant instabilities \citep[thus called ``self-confinement", e.g.,][]{Kulsrud1969,Wentzel1974}, since the external turbulence cascades to very low amplitudes at the small resonant scales of low-energy CRs \citep{Blasi2012,Evoli18}. In the self-confinement scenario, the most widely studied and often dominant gyro-resonant instability is the CR streaming instability (CRSI): large-scale CR pressure gradients
drive collective CR bulk motion (i.e., streaming) in the direction $- \vec{b} \vec{b}\cdot\nabla P_{\rm cr}$, which excites Alfv\'en waves when the drift velocity exceeds $v_A$. Within the reference frame of waves, the CRs are resonantly scattered towards isotropy, effectively locking their bulk flow to the local Alfv\'en speed if the wave amplitude (as limited by damping) is sufficiently high. By scattering CRs and damping with the background plasma, MHD waves facilitate energy and momentum transfer between CRs and the MHD gas. MHD-PIC simulations have extensively investigated the growth of CRSI, the role of wave damping, and the resulting saturation of the instability \citep{Bai2019,Bai2022,Plotnikov2021,Bambic_2021,Lemmerz2024}.


To test the paradigm of galactic-scale CR self-confinement with CRSI, \citet{Armillotta_2021} implemented a subgrid model for CR scattering, in which amplitudes are set by locally balancing CRSI wave excitation with ion-neutral and nonlinear Landau wave damping \citep[see also ][for similar implementations]{Hopkins_2021,Sike_2025, Thomas_2025}, coupled with the two-moment CR solver of \citet{JiangOh2018}. This framework was applied to study transport of GeV CR protons in realistic multiphase gas and magnetic fields, as obtained from high-resolution star-forming ISM simulations for both solar neighborhood and other environmental conditions \citep{Armillotta2022}. Extending the model to multiple energy groups and including CR electrons, \citet{Linzer_2025,Armillotta_2025} showed that both CR proton and electron spectra match observations across $1-100$GeV, with energy-dependent grammage also satisfying observational constraints. This initial success highlights the promise of two-moment frameworks equipped with subgrid CR transport, and motivates advanced CR–MHD methods that include gyrokinetic effects beyond CRSI along with more robust numerical schemes.

Notably, the CR pressure anisotropy instability (CRPAI) has been identified as a significant alternative process that drives CR self-confinement (first described by \cite{Kulsrud04} and further studied by \cite{Lazarian2006,Yan2011,Lebiga18,Zweibel20,Sun2024}). In this scenario, an initially isotropic CR distribution in phase space becomes prolate/oblate due to adiabatic invariance as CR fluid elements experience rapidly increasing/decreasing magnetic fields. The resulting CR pressure anisotropy then supplies free energy for wave growth. Distinct from the CRSI, the CRPAI generates both forward and backward propagating Alfv\'en waves with no preference for the $\nabla P_{\rm cr}$ direction. This characteristic makes CRPAI an important complement to the CRSI, especially at the confounding CR bottleneck structures where the CR and thermal pressure gradients are misaligned, flattening $\nabla P_{\rm cr}$ and deactivating the CRSI \citep{Wiener2017}. Consequently, the additional confinement provided by CRPAI could influence CR-cloud interactions \citep[e.g.,][]{Wiener2017,Bruggen2020} and regulate wind launching and mass loading in CR-overpressured regions \citep[e.g., bubbles, see][]{Schroer2022,Lancaster2024}. Also, CRPAI itself might dominate the wave excitations in environments with temporally or spatially rapidly changing magnetic fields, providing near-source heating \citep[e.g.,][]{Wagner2007,Caprioli2009}, modifying mixing layers in the multiphase ISM/CGM \citep{Ji2019,Drummond2023,Zhao2023} and affecting the thermal and hydrostatic stability of CGM halos \citep[e.g.,][]{Tsung2022}.

As uncertainties in CR feedback are intrinsically tied to unresolved transport physics, advancing CR feedback modeling to next level requires a general framework capable of incorporating all potential transport mechanisms. However, current two-moment schemes have limited extensibility. The formulation in \cite{JiangOh2018} implicitly assumes that Alfvén waves can only propagate along one direction (either along or opposed to the local $\vec{b}$). 
This is well motivated 
by CRSI, 
but is not consistent with 
more general Alfv\'en wave configurations as in CRPAI. The \cite{Thomas19} formulation in principle does allow general wave configurations, but their derivation only targets drift anisotropy, and incorporating additional physics would lead to further complications, especially in dealing with the wave subsystem.

In this work, we take a first-principles starting point similar to \cite{Thomas19}, and formulate a two-moment description for CR hydrodynamics that accommodates an arbitrary configuration of Alfv\'en waves interacting with the CR ensemble.
We include CR pressure anisotropy along with its effects in both the scattering term and the advection term. Our formulation provides a more general framework for CR feedback modeling, which has the flexibility to incorporate various CR transport modes such as the extrinsic turbulence, self-confinement by the CRSI, the CRPAI and potentially their combination. We implement our two-moment CR formulation into the \textit{Athena}++ code \citep{Stone2020} as a fluid species coupled to the MHD gas. To ensure numerical robustness, we also calculate characteristic wave speeds for the CR-MHD system and apply them in our HLLE Riemann solver. Furthermore, we perform von-Neumann stability analysis and derive a time-step criterion.

This paper is organized as follows: \autoref{sec:2} derives the framework for our CR hydrodynamics from first principles, beginning with the CR kinetic equation. \autoref{sec:3} outlines the numerical implementation of our CR module in the \textit{Athena}++ code, detailing considerations such as wave speeds adopted in the HLLE solver, and time step criterion prescribed to ensure numerical stability. \autoref{sec:4} and \autoref{sec:5} present benchmark tests validating code performance. \autoref{sec:6} discusses our formulation in comparison with prior two-moment schemes, potential astrophysical applications, limitations as well as future extensions of our current implementation. A preliminary analysis of characteristic waves in the CR-MHD system is presented in \autoref{append:waves}, followed by a von Neumann analysis to obtain the stability condition in \autoref{append:timestep}. In \autoref{append:cylindricalCRPAI}, we provide a simple analytical derivation of CR energy density evolution in cylindrical coordinates with CR scattering solely subject to CRPAI, which serves as a benchmark for comparison with our numerical tests.

\section{CR hydrodynamics} \label{sec:2}
In this section, we derive the two-moment CR equations to be coupled with the MHD gas.
\subsection{CR Evolution in Phase Space}

Following the calculations by \cite{Skilling1975} with minor modifications, we start by deriving the time evolution of the CR distribution function. The collisionless Boltzmann equation for relativistic particles is:
\begin{equation}\label{Boltzmann}
    \frac{\partial f}{\partial t}+\vec{v}\cdot\nabla f+\nabla_p\cdot\left(f\frac{d\vec{p}}{dt}\right)=0
\end{equation}
where $f\equiv f(\vec{x},\vec{p},t)$ is the CR distribution and $\vec{v}$ is the CR particle velocity. The net force $d\vec{p}/dt$ includes the dominant Lorentz force and a remaining force due to the Alfv\'en waves which is treated as a scattering process. In a frame comoving with the MHD gas, we benefit from the fact that CRs' gyromotion is axisymmetric about field lines, which is in general not true in the lab frame where CRs are transported across field lines. For this reason, we transform \autoref{Boltzmann} into a frame that moves at $\vec{u}_\perp\equiv \vec{u}-\vec{u}\cdot\vec{b}\vec{b}$,
where $\vec{u}$ is the background gas velocity and $\vec{b}\equiv\vec{B}/B$ is the unit vector along magnetic field. This frame is as close to the lab frame as possible while preserving the gyrotropic symmetry, because in the direction perpendicular to the magnetic field, this frame has no relative motion with respect to the field lines.

In this reference frame, the CR distribution has no dependence on the gyrophase $\phi$, and we can write it as $\bar{f}(\vec{x}, p,\mu, t)$ where $p$ is the magnitude of particle momentum and $\mu$ is the pitch-angle cosine $(\mu\equiv\vec{p}\cdot\vec{b}/p)$. After averaging over $\phi$, \cite{Skilling1975} writes 
the evolution equation in the frame moving at $\vec{u}_\perp$ as his equation (5). 
We find this is equivalent to the following form\footnote{This result is consistent with equation (5.21) in \cite{Zank2014} and equation (29) in \cite{Thomas19}, except they ignored the $\partial \vec{b}/\partial t$ terms}:
\begin{equation}\label{CRevolution}
    \begin{aligned}
    &\frac{\partial \bar{f}}{\partial t}+(\vec{u}_\perp+\mu v\vec{b})\cdot \nabla\bar{f}\\
    &+\left[\frac{1-3\mu^2}{2}(\vec{b}\vec{b}:\nabla\vec{u}_\perp)-\frac{1-\mu^2}{2}\nabla\cdot\vec{u}_\perp+\frac{\mu}{v}\vec{u}_\perp\cdot\frac{\partial \vec{b}}{\partial t}\right]p\frac{\partial \bar{f}}{\partial p}\\
    &+\left[v\nabla\cdot\vec{b}+\mu\nabla\cdot\vec{u}_\perp-3\mu(\vec{b}\vec{b}:\nabla\vec{u}_\perp)+\frac{3}{c}\vec{u}_\perp\cdot\frac{\partial \vec{b}}{\partial t}\right]\frac{1-\mu^2}{2}\frac{\partial \bar{f}}{\partial \mu}\\
    &=\left.\frac{\delta f}{\delta t}\right|_{\rm scatt}.
    \end{aligned}
\end{equation}

While the dominant Lorentz force has no net contribution after being averaged over $\phi$, 
effects due to small-amplitude Alfv\'en waves are incorporated in a scattering term on the right hand side (RHS). Usually this term is computed as a pure pitch-angle scattering in the frame of the wave, given by
\citep{Skilling1971}:
\begin{equation}
    \left.\frac{\delta f}{\delta t}\right|_{\rm scatt,\rm wave}=\frac{\partial }{\partial \mu}\left[\frac{1-\mu^2}{2}\nu(p,\mu)\frac{\partial f}{\partial \mu}\right]_{\rm wave}.
\end{equation}
Since particle energy is conserved in the wave frame, the scattering process becomes a pitch angle diffusion which is encoded by the scattering frequency $\nu(p,\mu)$. 

However, there is no longer an unitary wave frame if we consider the situation where both forward and backward propagating Alfv\'en waves are present, and this is another reason that we chose 
the $\vec{u}_\perp$ frame (instead of any wave frame) in the first place. In our frame, taking the scatterings by forward and backward waves mutually uncorrelated in phase, \cite{Skilling1975} has written down the scattering term as:
\begin{equation}\label{scattering}
    \left.\frac{\delta f}{\delta t}\right|_{\rm scatt}=\sum_{i=\pm}\frac{\gamma_i}{\gamma}\frac{\partial}{\partial \mu_i}\left[\frac{1-\mu_i^2}{2}\nu_i(p,\mu)\frac{\partial f_i}{\partial \mu_i}\right]
\end{equation}
with the transformation relations:
\begin{equation}
    \frac{\gamma_\pm}{\gamma}=1-\frac{\mu v}{c^2}\vec{b}\cdot\vec{w}_\pm,
\end{equation}
\begin{equation}
    \frac{\partial }{\partial \mu_\pm}=\left(1-\frac{2\mu \vec{b}\cdot\vec{w}_\pm}{v}\right)\frac{\partial}{\partial \mu}+\gamma m\vec{b}\cdot\vec{w}_\pm \frac{\partial }{\partial p}.
\end{equation}
Here, $\gamma$ is the Doppler factor of the particle and subscripts $+/-$ denote quantities as measured in the forward/backward wave frames, whose velocities are $\vec{w}_{\pm}\equiv \vec{u}\pm \vec{v}_A$ with $\vec{v}_A=\vec{B}/\sqrt{4\pi\rho}$ being the Alfv\'en velocity, and $\nu_+$ and $\nu_-$ are the scattering frequencies due to forward and backward propagating Alfv\'en waves. All the equations above are valid at least to $\mathcal{O}(w_\pm/c)$ where $c$ is the speed of light.

By keeping only the dominant terms in \autoref{scattering}, we can rewrite the scattering term into the Fokker-Plank form\footnote{Expanding \autoref{scattering} in fact gives $\sum_{i=\pm}\partial \left[\left(1-3\mu\vec{b}\cdot\vec{w}_\pm/c \right)F_{\mu,i}\right]/\partial \mu+{p^{-2}}\partial (p^2F_{p,i})/\partial p$. However, when we take moments, the $3\mu\vec{b}\cdot\vec{w}_\pm/c$ term is always much less than order unity and we can safely omit that term.}:
\begin{equation}\label{FPscatter}
    \left.\frac{\delta f}{\delta t}\right|_{\rm scatt}=\sum_{i=\pm}\left[\frac{\partial F_{\mu,i}}{\partial \mu}+\frac{1}{p^2}\frac{\partial}{\partial p}\left(p^2F_{p,i}\right)\right],
\end{equation}
where 
\begin{equation}
    F_{\mu,\pm}\equiv\frac{1-\mu^2}{2}\nu_\pm(p,\mu)\left(\frac{\partial \bar{f}}{\partial \mu}+p\frac{\vec{b}\cdot\vec{w}_\pm}{c}\frac{\partial \bar{f}}{\partial p}\right),
\end{equation}
\begin{equation}
F_{p,\pm}\equiv p\frac{\vec{b}\cdot\vec{w}_\pm}{c}F_{\mu,\pm}.
\end{equation}
\autoref{CRevolution} and \autoref{FPscatter} together lay out the foundation upon which we build our hydrodynamic equations for CRs.

\subsection{CR Fluid Dynamics}
In this subsection we first introduce some definitions of our method, then formulate the CR fluid equations by taking the traditional moment hierarchy. From here we omit the spatial coordinate in the CR distribution function since we will mostly focus on integrals over momentum space which are spatially local.

To establish a two-moment description, three moments are involved: CR energy density $\mathcal{E}_{\rm cr}$, energy flux $\vec{F}_{\rm cr}$ and pressure tensor $\mathbf{P}_{\rm cr}$, which are defined as:
\begin{equation}\label{eq:Ecrdef}
    \mathcal{E}_{\rm cr}=\int E(p)p^2 \bar{f}dpd\mu d\phi,
\end{equation}
\begin{equation}\label{eq:Fcrdef}
    \vec{F}_{\rm cr}=\int \vec{v}E(p)p^2 \bar{f}dpd\mu d\phi,
\end{equation}
\begin{equation}\label{eq:Pcrdef}
    \mathbf{P}_{\rm cr}=\int \vec{p}\vec{v}p^2 \bar{f}dpd\mu d\phi,
\end{equation}
where $E(p)$ is the energy of a CR particle with momentum $p$. In the derivation below we estimate it as $E(p)\approx pc$ in the ultra-relativistic limit where $v\approx c$.

Given that $\bar{f}$ is gyrotropic, we can do the $\phi$ integral first and find $\vec{F}_{\rm cr}$ only has a component parallel to $\vec{B}$,
\begin{equation}\label{eq:Fparalleldef}
    \vec{F}_{\rm cr}=\vec{b}\int \mu v E(p)p^2 \bar{f}dpd\mu\equiv F_{\rm cr,\parallel}\vec{b},
\end{equation}
while $\mathbf{P}_{\rm cr}$ is diagonal in a local orthogonal coordinate system where $\hat{x}\parallel\vec{b}$:
\begin{equation}\label{eq:Pcrmatrixdef}
    \mathbf{P}_{\rm cr}=\begin{pmatrix}

    P_{\rm cr,\parallel} & 0 & 0\\
    0 & P_{\rm cr,\perp} & 0\\
    0 & 0 & P_{\rm cr,\perp}
    
    \end{pmatrix},
\end{equation}
with

\begin{equation}\label{eq:Pparalleldef}
    P_{\rm cr,\parallel}=\int \mu^2 pv p^2\bar{f}dpd\mu
\end{equation}

\begin{equation}\label{eq:Pperpdef}
    P_{\rm cr,\perp}=\int \frac{1-\mu^2}{2} pv p^2\bar{f}dpd\mu
\end{equation}
respectively being the pressure components parallel and perpendicular to the magnetic field, such that ${\rm tr}(\mathbf{P}_{\rm cr})=\mathcal{E}_{\rm cr}$. For convenience, we use the shorthand $P_{\rm cr}\equiv \mathcal{E}_{\rm cr}/3$ to denote the average of the diagonal terms of $\mathbf{P}_{\rm cr}$ in the rest of this paper.

The CR pressure anisotropy $\Delta P_{\rm cr}$ is then given by 
\begin{equation}\label{eq:DeltaPcrdef}
    \Delta P_{\rm cr}\equiv P_{\rm cr,\parallel}-P_{\rm cr,\perp}=\int P_2(\mu) pvp^2\bar{f}dpd\mu
\end{equation}
where $P_2 \equiv(3 \mu^2-1)/2$, with $P_n(\mu)$ denoting the $n$-th order Legendre polynomial. In the expressions above and after we omit the $2\pi$ factor from the $d\phi$ integral, as it appears in all terms and thus eventually cancels in the fluid equations.

Next, as any fluid description, we need a closure for the system. A commonly used isotropic closure for CRs is
$\mathbf{P}_{\rm cr}=\mathcal{E}_{\rm cr}\mathbf{I}/3$, where $\mathbf{I}$ is the unit tensor. Physically, this closure is motivated by the observational fact that the CR distribution $\bar{f}(p,\mu,t)$
in our Galaxy only has a very low level of anisotropy of order $\mathcal{O}(10^{-4})$ in the pitch angle $\mu$ \citep{Kulsrud04}, i.e. $\bar{f}(p,\mu,t)\approx f_0(p,t)$. If we use such an isotropic ansatz, or even add a first-order component representing a drift anisotropy (i.e., $\bar{f}(p,\mu,t)=f_0(p,t)+f_1(p,t)P_1(\mu)$ for $P_1(\mu)\equiv\mu$, where the isotropic component dominates $f_0\gg f_1$), then by directly plugging it back into the definitions \autoref{eq:Ecrdef} and \autoref{eq:Pcrmatrixdef}, we see $P_{\rm cr,\parallel}=P_{\rm cr,\perp}=\mathcal{E}_{\rm cr}/3$.

Even if the level of anisotropy is quite small, $\Delta P_{\rm cr}$ can play an important role in both macro- and microphysical processes. 
Microscopically, the CR pressure anisotropy can trigger gyroresonant instability \citep{Kulsrud04,Yan2011,Lazarian2006,Sun2024} similar to the CRSI driven by a drift anisotropy, and thus bring in a new channel of self-confinement. This mechanism can potentially have many macroscopic consequences. For example in the case of CR bottlenecks \citep[e.g,][]{Skilling1971,Wiener2017} where the $\mathcal{E}_{\rm cr}$ gradient is flattened and the CRSI is deactivated, the CRPAI may then dominate and govern CR momentum transfer to the gas, which may alter the properties of CR-driven wind. In situations such as a supernova remnant where the system undergoes rapid expansion,
the CR pressure anisotropy could be efficiently driven and induce a high level of heating \citep{Zweibel20}. Additionally, consider a magnetic field with a focusing structure and suppose there is no MHD wave scattering at all, then CR flux should still be affected by the $\Delta P_{\rm cr}$ in the advection term $\nabla\cdot\mathbf{P}_{\rm cr}$ due to adiabatic invariance.

Given the potentially rich physics related with $\Delta P_{\rm cr}$, we would like to generalize our CR formulation to self-consistently incorporate CR pressure anisotropy,
both in the advection and source terms, together with the scattering term due to CRPAI.
This is done by adopting an \textit{ansatz} for the CR distribution $\bar{f}$: 
\begin{equation}\label{ansatz}
    \bar{f}(p,\mu,t)=f_{0}(p,t)+f_1(p,t)P_1(\mu)+f_2(p,t)P_2({\mu}).
\end{equation}
Expanding CR distributions $f$ by Legendre polynomials is canonical in the study about CR transport because of their natural geometric characteristics in pitch angle \citep[e.g.,][]{Klimas1971,Earl1973,Webb1987,Zank2000,Snodin2006,Litvinenko2013,Rodrigues2019,Thomas19,Zweibel20}. Mathematically, a rigorous expansion certainly must involve a complete set of Legendre polynomials as basis functions. However, from the definitions in \autoref{eq:Ecrdef}, \autoref{eq:Fparalleldef}, \autoref{eq:DeltaPcrdef} and the relation ${\rm tr}(\mathbf{P}_{\rm cr})=\mathcal{E}_{\rm cr}$, we see Legendre polynomials up to the second order are already sufficient to incorporate $\Delta P_{\rm cr}$.
We thus truncate at second order. 
Given the orthogonality of Legendre polynomials, each $f_n(p,t)\ (n=0,1,2)$ component separately corresponds to a fluid moment:

\begin{equation}\label{eq:Ecrf0}
    \mathcal{E}_{\rm cr}=2\int_0^{+\infty} E(p)p^2 f_0(p)dp,
\end{equation}

\begin{equation}\label{eq:Fcrf1}
    F_{\rm cr,\parallel}=\frac{2c}{3}\int_0^{+\infty} E(p)p^2 f_1(p)dp,
\end{equation}

\begin{equation}\label{eq:DeltaPcrf2}
    \Delta P_{\rm cr}=\frac{2}{5}\int_0^{+\infty} E(p)p^2 f_2(p)dp.
\end{equation}

With these conventions, we now take the $\mathcal{E}_{\rm cr}$ and $F_{\rm cr,\parallel}/c^2$ moments on top of the kinetic \autoref{CRevolution}, yielding left hand sides (LHSs) as:
\begin{equation}\label{EcrLHS}
    \frac{\partial \mathcal{E}_{\rm cr}}{\partial t}+\nabla\cdot\left(\vec{F}_{\rm cr}+\mathcal{E}_{\rm cr}\vec{u}_\perp+\vec{u}_\perp\cdot\mathbf{P}_{\rm cr}\right)-\vec{u}_\perp \cdot(\nabla\cdot \mathbf{P}_{\rm cr})-\frac{\vec{u}_\perp}{c^2}\cdot\frac{\partial \vec{F}_{\rm cr}}{\partial t}
\end{equation}
and
\begin{equation}\label{FcrLHS}
    \vec{b}\cdot\left\{\frac{\partial [\vec{F}_{\rm cr}+(\mathcal{E}_{\rm cr}+P_{\rm cr,\perp})\vec{u}_\perp]}{c^2\partial t}+\nabla\cdot\left(\mathbf{P}_{\rm cr}+\frac{\vec{F}_{\rm cr}\vec{u}_\perp+\vec{u}_\perp \vec{F}_{\rm cr}}{c^2}\right)\right\}.
\end{equation}
We will derive the RHSs associated with moment integrals of the scattering terms after introducing relevant microphysics in the next subsection. Note that the fluid quantities $\mathcal{E}_{\rm cr}$, $\vec{F}_{\rm cr}$ and $\mathbf{P}_{\rm cr}$ here are still in the $\vec{u}_\perp$ frame, and in later section we will transform these equations into the lab frame.
The exact closure resulting from our \textit{ansatz} is given by \autoref{eq:closure} in a later section.

\subsection{CR Scattering}
Computing moment integrals of the scattering terms (\autoref{FPscatter}) involves uncertainties because of our limited knowledge about the scattering rates $\nu_\pm (p,\mu)$. In this subsection, we discuss microphysics of CR scattering, as well as our treatments of the scattering terms.

CRs interact primarily with Alfvén waves through resonant scattering, which governs their transport and coupling to the thermal plasma. Assuming a circularly polarized Alfv\'en wave with a wave vector $\vec{k}$ and a Doppler-shifted angular frequency $\omega$, then a particle can resonantly scatter off such a wave if the following gyro-resonant condition is satisfied \citep{Schlickeiser02,Kulsrud04}:

\begin{equation}\label{resonance_condition}
    k_\parallel \mu v-\omega=\pm \frac{n\Omega_{\rm cr}}{\gamma}
\end{equation}
where $k_\parallel\equiv\vec{k}\cdot\vec{b}$ is the parallel component, and $\Omega_{\rm cr}\equiv eB/mc$ is the CR particle cyclotron frequency (with $e$ being particle charge) so that $\Omega_{\rm cr}/\gamma$ becomes the relativistic gyrofrequency. Here $n$ is a natural number representing the order of the resonance. While $n>1$ describes more general cases where $\vec{k}$ can be oblique to the magnetic field, in our current formulation we restrict to the 1D case with waves parallel to magnetic fields $(k_{\parallel}=k)$, motivated in part by the finding of \cite{Zeng2025} that 2D effects are relatively minor. We choose the convention where the $+/-$ signs on the RHS separately denote right-handed/left-handed polarized waves.

In the comoving frame, the resonance condition \eqref{resonance_condition} simplifies to $k(\mu v- v_A)=\pm\Omega_{\rm cr}/\gamma$, describing a phase match between the particle's gyromotion and the Alfvén wave. Only resonance enables efficient scattering, otherwise, non-resonant CRs experience rapidly fluctuating fields that yield negligible net interaction. For relativistic CRs with $\mu v\gg v_A$, the condition further reduces to $k\mu v\approx \pm \Omega_{\rm cr}/\gamma$.


Restricting to waves propagating parallel to the background magnetic field, there are generally four wave modes characterized by both direction and polarization. Accordingly, we subdivide scattering rates into $\nu_+^L$, $\nu_+^R$, $\nu_-^L$ and $\nu_-^R$ where subscripts $\pm$ denote directions along or opposite to magnetic fields, while superscripts $L,R$ denote polarization, and define
\begin{equation}
    \nu_+(p,\mu)=\nu_+^L(p,\mu)+\nu_+^R(p,\mu),
\end{equation}
\begin{equation}
    \nu_-(p,\mu)=\nu_-^L(p,\mu)+\nu_-^R(p,\mu).
\end{equation}
The values of these terms depend on the intensities of the corresponding resonant waves \citep{Schlickeiser1989}, which are determined by specific physical processes. However, even without a specified wave-driving mechanism, from the resonant condition $k\mu v\approx \pm \Omega_{\rm cr}/\gamma$ we can see forward/backward traveling particles $(\mu>0/\mu<0)$ only resonantly interact with right-hand/left-hand polarized waves $(+/-\ {\rm in\ front\ of\ }\Omega_{\rm cr})$, such that:
\begin{equation}
    \nu_\pm^L (p,\mu)
    \begin{cases}
      =0, & \mu>0 \\
      >0, & \mu<0
    \end{cases}
\end{equation}
\begin{equation}
    \nu_\pm^R (p,\mu)
    \begin{cases}
      >0, & \mu>0 \\
      =0, & \mu<0
    \end{cases}  .  
\end{equation}

We now proceed with taking the $\mathcal{E}_{\rm cr}$ and $F_{\rm cr,\parallel}$ moments of \autoref{FPscatter}. By firstly integrating over $\mu$ by parts and then expanding $\partial \bar{f}/\partial \mu$ using our ansatz in \autoref{ansatz}, we can combine all the dependence on $\mu$ with $\nu_\pm^{L,R}(p,\mu)$, and define momentum-wise effective scattering rates:
\begin{equation}
\begin{aligned}
\nu_{\pm,\rm eff}^{L,R}(p)&\equiv\frac{\int_{-1}^1 (1-\mu^2)\nu_{\pm}^{L,R}(p,\mu)/2d\mu}{\int_{-1}^1(1-\mu^2)/2d\mu}\\
&=\frac{3}{2}\int_{-1}^1\frac{1-\mu^2}{2}\nu_\pm^{L,R}(p,\mu)d\mu
\end{aligned}
\end{equation}

Higher order Legendre polynomial $P_2(\mu)$ associated with $\Delta P_{\rm cr}$ introduces an additional $\mu$ under the integral, whose effect is to change sign in front of the effective scattering rates and suppress the magnitude. We thus use a factor $0<\alpha<1$ to account for this effect, such that:

\begin{equation}
\frac{3}{2}\int_{-1}^1\mu\frac{1-\mu^2}{2}\nu_\pm^Ld\mu=\frac{3}{2}\int_{-1}^0\mu\frac{1-\mu^2}{2}\nu_\pm^Ld\mu\equiv-\alpha\nu_{\pm,\rm eff}^L
\end{equation}

\begin{equation}
\frac{3}{2}\int_{-1}^1\mu\frac{1-\mu^2}{2}\nu_\pm^Rd\mu=\frac{3}{2}\int_{0}^1\mu\frac{1-\mu^2}{2}\nu_\pm^Rd\mu\equiv\alpha\nu_{\pm,\rm eff}^R
\end{equation}
The exact value of the factor $\alpha$ can be calibrated from MHD-PIC simulations \citep[e.g.,][]{Bai2022,Sun2023}.

Finally, for CR hydrodynamics we define momentum-integrated effective scattering rates by averaging $\nu_{\pm,\rm eff}^{L,R}(p)$ weighted by fluid moments:

\begin{equation}
    \widetilde{\nu}_{\pm}^{L,R}\equiv\frac{\int\nu_{\pm,\rm eff}^{L,R}(p)p^2 E(p)f_m(p)dp}{\int p^2 E(p)f_m(p)dp}
\end{equation}
where $m=0,1,2$. Strictly speaking, the overall effective scattering rates $\widetilde{\nu}_\pm^{L,R}$ have a dependence on the weighting function $f_m(p)$, but since we are mostly interested in GeV CRs within a rather narrow energy band, the resulting differences are expected to be minor. 
Additionally, MHD-PIC simulations can be used to calibrate such differences.  

Using all the notations above, we list the moment integrals of the scattering terms that separately correspond to the RHS of the $\mathcal{E}_{\rm cr}$ and $F_{\rm cr,\parallel}$ equations:

\begin{equation}\label{eq:EcrRHS}
    \begin{aligned}
    &\mathcal{E}_{\rm cr}\ {\rm RHS}=-\sigma_+\left[\vec{F}_{\rm cr}-(\mathcal{E}_{\rm cr}+P_{\rm cr})\vec{w}_+\right]\cdot\vec{w}_{+,\parallel} \\
    &-\sigma_-\left[\vec{F}_{\rm cr}-(\mathcal{E}_{\rm cr}+P_{\rm cr})\vec{w}_-\right]\cdot\vec{w}_{-,\parallel} \\
    &+5\alpha c[(\sigma_+^L-\sigma_+^R)w_{+,\parallel}+(\sigma_-^L-\sigma_-^R)w_{-,\parallel}]\Delta P_{\rm cr}
    \end{aligned}
\end{equation}

\begin{equation}\label{eq:FcrRHS}
    \begin{aligned}
    &F_{\rm cr,\parallel}/c^2\ {\rm RHS}=-\sigma_+\left[\vec{F}_{\rm cr}-(\mathcal{E}_{\rm cr}+P_{\rm cr})\vec{w}_+\right]\cdot\vec{b}\\
    &-\sigma_-\left[\vec{F}_{\rm cr}-(\mathcal{E}_{\rm cr}+P_{\rm cr})\vec{w}_-\right]\cdot\vec{b}\\
    &+5\alpha c\left(\sigma_+^L-\sigma_+^R+\sigma_-^L-\sigma_-^R\right)\Delta P_{\rm cr}
    \end{aligned}
\end{equation}
where $\vec{w}_{\pm,\parallel}\equiv\vec{w}_{\pm}\cdot\vec{b}\vec{b}$ is the parallel component. From here we define scattering coefficients $\sigma_\pm^{L,R}\equiv\widetilde{\nu}_\pm^{L,R}/c^2$ and use the shorthand $\sigma_\pm\equiv\sigma_\pm^L+\sigma_\pm^R$. In principle, $\alpha$ could be any positive number smaller than one, but from here we adopt $\alpha=1/5$ to simplify the form of our equations. Note that the origin of the Alfv\'en waves has not been specified yet, and our formulation remains valid so long as CR transport arises from resonant scattering with the waves.

\subsection{CR Hydrodynamic Equations in Lab Frame}

Equating LHS \autoref{EcrLHS} and \autoref{FcrLHS} with RHS \autoref{eq:EcrRHS} and \autoref{eq:FcrRHS} gives CR fluid equations in the $\vec{u}_\perp$ frame, but in the end we need equations in the lab frame. In doing so, we first need the transformation relations for CR fluid quantities $\mathcal{E}_{\rm cr}$, $\vec{F}_{\rm cr}$ and $\mathbf{P}_{\rm cr}$, and then derive the perpendicular component of $\vec{F}_{\rm cr}$ arising from advection across field lines.

As CRs are relativistic particles, CR hydrodynamics shares similarities to RHD particularly in frame transformation relations. \cite{Thomas19} showed the results for CRs by combining $\mathcal{E}_{\rm cr}$, $\vec{F}_{\rm cr}$ and $\mathbf{P}_{\rm cr}$ as a rank-2 energy momentum tensor and then applying a Lorentz transformation. We slightly extended their results to the case with an anisotropic $\mathbf{P}_{\rm cr}$, and list the relations below. Our calculation and final results are in direct analogy with the relations in RHD \citep[e.g.,][]{Mihalas84,Castor07,Jiang2012}:

\begin{equation}
    \mathcal{E}_{\rm cr,lab}=\mathcal{E}_{\rm cr}+\frac{2\vec{u}_\perp \cdot\vec{F}_{\rm cr}}{c^2}+\mathcal{O}\left(\frac{u_\perp^2}{c^2}\right)
\end{equation}

\begin{equation}
    \vec{F}_{\rm cr,lab}=\vec{F}_{\rm cr}+\mathcal{E}_{\rm cr}\vec{u}_\perp+\vec{u}_\perp\cdot\mathbf{P}_{\rm cr}+\mathcal{O}\left(\frac{u_\perp^2}{c^2}\right)
\end{equation}

\begin{equation}
    \mathbf{P}_{\rm cr,lab}=\mathbf{P}_{\rm cr}+\frac{\vec{F}_{\rm cr}\vec{u}_\perp+\vec{u}_\perp\vec{F}_{\rm cr}}{c^2}+\mathcal{O}\left(\frac{u_\perp^2}{c^2}\right)
\end{equation}
Neglecting $\mathcal{O}(u_\perp^2/c^2)$ terms, we see $\mathcal{E}_{\rm cr,lab}=\mathcal{E}_{\rm cr}$ since $\vec{u}_\perp \perp \vec{F}_{\rm cr}$, thus we do not distinguish between them. It is also evident that $\vec{F}_{\rm cr,lab}\cdot\vec{b}=F_{\rm cr,\parallel}$, i.e., the parallel components in the lab frame and the $\vec{u}_\perp$ frame are identical.

One issue associated with transforming $\mathbf{P}_{\rm cr}$ to the lab frame is the difficulty of determining $\Delta P_{\rm cr}$ in the lab frame. In fact it makes even no sense to define $\Delta P_{\rm cr}$ in the lab frame because the CR distribution $f$ is no longer symmetric about the gyro phase.
However, quasi-linear theory (QLT) suggests 
that the anisotropic component in the ansatz \autoref{ansatz} will approximately satisfy $|f_1/f_0|\sim v_A/c$ and $|f_2/f_0|\sim v_A/c$ \citep{Sun2024}, which by equations \autoref{eq:Ecrf0}-\autoref{eq:DeltaPcrf2} implies $|\vec{F}_{\rm cr}|\sim v_A \mathcal{E}_{\rm cr}$ and $|\Delta P_{\rm cr}|\sim v_A \mathcal{E}_{\rm cr}/c$ (see also Section \autoref{subsec:subgrid}).
Consequently, the additional term arising from $\mathbf{P}_{\rm cr}$ transformation is $\sim \mathcal{O}(v_A u_\perp/c^2)$, which is generally significantly smaller than $\Delta P_{\rm cr}$. We can thus consider $\mathbf{P}_{\rm cr,lab}\approx\mathbf{P}_{\rm cr}$, and regard $\Delta P_{\rm cr}$ somewhat as a scalar. These considerations imply that the closure for $\mathbf{P}_{\rm cr,lab}$ becomes
\begin{equation}\label{eq:closure}
    \mathbf{P}_{\rm cr,lab}\approx\mathbf{P}_{\rm cr}=\left(\frac{\mathcal{E}_{\rm cr}-\Delta P_{\rm cr}}{3}\right)\mathbf{I}+\Delta P_{\rm cr}\vec{b}\vec{b}
\end{equation}
where $\Delta P_{\rm cr}$ is prescribed by \autoref{eq:DeltaPcrpres} in the next subsection.

With the frame transformation relations at hand, in principle we can already solve for the CR evolution as a fluid. However, solving $F_{\rm cr,\parallel}$ is inconvenient because the ``parallel" direction is not uniform and is evolving.
To remedy this, we introduce the perpendicular force whose information was lost during gyro-averaging, so that we can evolve $\vec{F}_{\rm cr,lab}$ instead of $F_{\rm cr,\parallel}$.
In doing so, we note that fluid motion and CR advection can cause a non-zero perpendicular CR flux, which induces a perpendicular CR current $\vec{j}_{\rm cr,\perp}=e\vec{F}_{\rm cr,\perp}/\left<\gamma\right>mc^2$ where $\left<\gamma\right>$ is the mean gamma factors of CR particles. The Lorentz force due to such a large-scale CR current is\footnote{In the eventual square bracket we use the complete vectors instead of perpendicular components, because it is numerically convenient and the parallel component has no contribution.}

\begin{equation}\label{Lorentz}
\begin{aligned}
    \frac{\vec{j}_{\rm cr,\perp}\times\vec{B}}{c}&=\frac{\Omega_{\rm cr}}{\left<\gamma\right>c^2}\vec{F}_{\rm cr,\perp}\times\vec{b}\\
    &=\frac{\Omega_{\rm cr}}{\left<\gamma\right>c^2}\left[\vec{F}_{\rm cr,lab}-(\mathcal{E}_{\rm cr}+P_{\rm cr})\vec{u}\right]\times\vec{b}
    \end{aligned}
\end{equation}

\cite{Thomas19} suggested this Lorentz force should counteract the perpendicular CR inertia and establish a dynamical equilibrium $\vec{j}_{\rm cr}\times\vec{B}/c=\nabla_\perp P_{\rm cr}$ on a kinetic time scale, which keeps the $\vec{F}_{\rm cr}$ in the comoving frame parallel to the magnetic field. Given such a consideration and the fact that \autoref{FcrLHS} is obviously the parallel component of $c^{-2}\partial \vec{F}_{\rm cr,lab}/\partial t+\nabla\cdot\mathbf{P}_{\rm cr,lab}$, we expect the additional perpendicular component to balance the Lorentz force, and thus write down the full CR flux equation as:

\begin{equation}
\begin{aligned}
    &\frac{1}{c^2}\frac{\partial \vec{F}_{\rm cr,lab}}{\partial t}+\nabla\cdot\mathbf{P}_{\rm cr,lab}=\\
    &-\sigma_+\left[\vec{F}_{\rm cr,lab}-(\mathcal{E}_{\rm cr}+P_{\rm cr})(\vec{u}+\vec{v}_A)\right]\cdot\vec{b}\vec{b}\\
    &-\sigma_-\left[\vec{F}_{\rm cr,lab}-(\mathcal{E}_{\rm cr}+P_{\rm cr})(\vec{u}-\vec{v}_A)\right]\cdot\vec{b}\vec{b}\\
    &+\left(\sigma_+^L-\sigma_+^R+\sigma_-^L-\sigma_-^R\right)c\Delta P_{\rm cr}\vec{b}\\
    &+\frac{\Omega_{\rm cr}}{\left<\gamma\right>c^2}\left[\vec{F}_{\rm cr,lab}-(\mathcal{E}_{\rm cr}+P_{\rm cr})\vec{u}\right]\times\vec{b}
    \end{aligned}
\end{equation}
$\Omega_{\rm cr}$ is typically a very large parameter and hence numerically keeps the CR flux aligned with magnetic fields in the comoving frame.

Next, we can dot the CR flux equation above with $\vec{u}_\perp$ to eliminate the $\partial\vec{F}_{\rm cr}/\partial t$ term in \autoref{EcrLHS}, and reach the following CR energy equation in the lab frame\footnote{When we dot $\vec{u}_\perp$ with $c^{-2}\partial \vec{F}_{\rm cr}/\partial t$, we expect the perpendicular components are in dynamic equilibrium and therefore still assume $\vec{F}_{\rm cr}\parallel\vec{b}$. Then the additional terms arising from frame transformation only give out $\mathcal{O}(u_\perp^2/c^2)$ terms which we neglect. Basically, we have $\vec{u}_\perp\cdot(c^{-2}\partial\vec{F}_{\rm cr}/\partial t+\nabla\cdot\mathbf{P}_{\rm cr})=\vec{u}_\perp\cdot(\vec{j}_{\rm cr,\perp}\times\vec{B})/c$.}:

\begin{equation}
    \begin{aligned}
    &\frac{\partial \mathcal{E}_{\rm cr}}{\partial t}+\nabla\cdot\vec{F}_{\rm cr,lab}=\\
    &-\sigma_+\left[\vec{F}_{\rm cr,lab}-(\mathcal{E}_{\rm cr}+P_{\rm cr})(\vec{u}+\vec{v}_A)\right]\cdot(\vec{u}_{\parallel}+\vec{v}_A)\\
    &-\sigma_-\left[\vec{F}_{\rm cr,lab}-(\mathcal{E}_{\rm cr}+P_{\rm cr})(\vec{u}-\vec{v}_A)\right]\cdot(\vec{u}_{\parallel}-\vec{v}_A)\\
    &+\left[(\sigma_+^L-\sigma_+^R)(u_\parallel+v_A)+(\sigma_-^L-\sigma_-^R)(u_\parallel-v_A)\right]c\Delta P_{\rm cr}\\
    &+\frac{\Omega_{\rm cr}}{\left<\gamma\right>c^2}\left(\vec{F}_{\rm cr,lab}\times\vec{b}\right)\cdot\vec{u}
    \end{aligned}
\end{equation}
This concludes our derivation of the new CR hydrodynamic equations that properly take into account CR pressure anisotropy.

\subsection{Subgrid Physics}\label{subsec:subgrid}

Our new CR formulation, while more complete, relies on additional input as subgrid physics. The primary component is the scattering coefficients $\sigma_{\pm}^{L,R}$, which essentially encapsulate most of the necessary microphysics, which needs to be specified by the user. Here we briefly discuss several typical 
scenarios below.

\begin{enumerate}
    \item Streaming Instability

    A large-scale CR pressure gradient $\nabla P_{\rm cr}$ drives CR bulk motion (streaming), which triggers CRSI once this bulk motion exceeds the Alfv\'en speed. The CRSI excites two branches of the Alfv\'en waves propagating down the direction of $\nabla P_{\rm cr}$ projected along the magnetic field \citep{Kulsrud1969,Wentzel1974}. 
    
    Assuming a steady balance between wave damping and CRSI-driven growth, the scattering coefficients have been shown to be \citep{Kulsrud04,Bai2019,Bai2022,Armillotta_2021}:
    \begin{equation}\label{eq:sigma_CRSI}
        \begin{aligned}
            &\sigma_+^L=\sigma_+^R\sim \frac{|\vec{b}\cdot\nabla P_{\rm cr}|}{v_A P_{\rm cr}}\frac{\Omega_{\rm cr}}{\nu_{\rm damp}}\frac{\rho_{\rm cr}}{\rho}\\
            &\sigma_-^L=\sigma_-^R=0
        \end{aligned}
    \end{equation}
    corresponding to the case $\nabla P_{\rm cr}\cdot\vec{B}<0$ where Alfv\'en waves along the magnetic field are driven. For $\nabla P_{\rm cr}\cdot\vec{B}>0$, the non-zero contribution instead lies in the $\sigma_-^{L,R}$ branch with the same magnitude. Here $\nu_{\rm damp}$ is the damping rate, and $\rho$ is the ionized gas density.

    With this prescription, the resulting equations are consistent with the current two-moment CR fluid formulation \citep[e.g.][]{JiangOh2018,Thomas19} as expected in the situation solely subject to the CRSI. The above relation applies for either linear or nonlinear wave damping (e.g. ion-neutral or nonlinear Landau; \citep{Kulsrud1969}). For ion-neutral damping, appropriate in neutral gas, the resulting $\sigma_{\pm}^{L,R}$ is linear in $|\nabla P_{\rm cr}|$;  for nonlinear Landau damping, appropriate in ionized gas, $\sigma_{\pm}^{L,R}$ varies as $\sqrt{|\nabla P_{\rm cr}|}$ \citep{Armillotta_2021}.

    \item Pressure Anisotropy Instability

    The CRPAI is a less studied version of the CR gyro-resonant instability, which can be triggered when the level of CR pressure anisotropy $\Delta P_{\rm cr}/P_{\rm cr}\gtrsim v_A/c$ \citep{Kulsrud04,Zweibel20}. It triggers one branch of Alfv\'en wave along each propagating direction. In reality, this may occur when the bulk CRs experience a varying magnetic field, e.g., when CRs travel through magnetic fields with large spatial variations, or the background field undergoes shear motion. Recent work \citep[e.g.,][]{Sun2024} has isolated this effect using an expanding/compressing box setup. They derived the scattering coefficients based on QLT, and calibrated the results using MHD-PIC simulations. The result is:

\begin{equation}\label{eq:CRPAIscattpres}
    \begin{aligned}
        &\sigma_+^L=\sigma_-^R\sim\frac{1}{v_A c}\left|\frac{\dot{B}}{B}\right|\frac{\Omega_{\rm cr}}{\nu_{\rm damp}}\frac{\rho_{\rm cr}}{\rho}\\
        &\sigma_+^R=\sigma_-^L=0
    \end{aligned}
\end{equation}

This applies for $\Delta P_{\rm cr}<0$ (increasing field strength). If instead $\Delta P_{\rm cr}>0$, then the roles reverse: $\sigma_+^L=\sigma_-^R=0$, while $\sigma_+^R=\sigma_-^L$ takes the non-zero value given above.

\item Extrinsic Turbulence

    If the MHD waves are driven by extrinsic MHD turbulence injection and the following energy cascade, then the waves have no preferred direction or polarization. We can thus assume $\sigma_+^L=\sigma_-^L=\sigma_+^R=\sigma_-^R\equiv\sigma_{\rm turb}/4$. By substituting this relation and use the steady-state $F_{\rm cr}$, we can obtain the equation describing pure diffusive transport of $\mathcal{E}_{\rm cr}$, which directly resembles previous results \citep[e.g., equation (30) in][]{Zweibel17}.
    
\end{enumerate}


The other ingredient for user input is the value of CR pressure anisotropy $\Delta P_{\rm cr}$, which is again associated with the CRPAI.
Similarly based on QLT, by balancing wave growth and (linear) damping, together with the growth of pressure anisotropy and quasi-linear diffusion, 
the steady state value of $\Delta P_{\rm cr}$
can be estimated as
\citep{Sun2024}:

\begin{equation}\label{eq:DeltaPcrpres}
    \frac{\Delta P_{\rm cr}}{P_{\rm cr}}\sim\pm\frac{v_A}{c}\frac{\nu_{\rm damp}}{\Omega_{\rm cr}}\frac{\rho}{\rho_{\rm cr}}
\end{equation}
The $+/-$ signs  correspond to a decreasing/increasing magnetic field strength that makes $P_{\rm cr,\perp}$ smaller/larger than $P_{\rm cr,\parallel}$. 

We highlight two important points behind the aforementioned prescriptions. First, the scattering coefficients and CR pressure anisotropy are derived assuming steady-state balance between wave driving and damping. Typically, this balance can be established on an instability growth timescale $\sim (n_{\rm ion}/n_{\rm cr})\Omega_{\rm cr}^{-1}\sim 10^{3-4}\ \rm yr$, which is orders of magnitude shorter than the dynamic timescale of the macroscopic system. Therefore, it usually suffices to adopt such steady-state estimates. Second, we note that these prescriptions were calculated by assuming either the CRSI or CRPAI being the only instability at work. In reality, the CRPAI and CRSI are likely to act together, although the outcome is not additive, because each of them amplify (different) two out of four parallel-propagating Alfv\'en waves, while damping the other two. For present purposes, 
we only consider the case with pure CRSI or CRPAI, while leaving the more general case to future work.

\section{Numerical Implementation}\label{sec:3}
We implement our CR equations in the \textit{Athena}++ MHD code \citep{Stone2020} as a CR fluid module, and in this section, we describe numerical details of our implementation.

For notational convention, all quantities without an explicit specifier are, hereafter, by default in lab frame.

\subsection{Governing Equations}\label{subsec:governeq}
We start by presenting the full equations that describe our system consisting of MHD gas and CR fluid:

\begin{equation}\label{mass_cons}
    \frac{\partial \rho}{\partial t}+\nabla\cdot(\rho\vec{u})=0
\end{equation}

\begin{equation}\label{momentum_cons}
\begin{aligned}
    \frac{\partial (\rho\vec{u})}{\partial t}+&\nabla\cdot(\rho \vec{u}\vec{u}-\vec{B}\vec{B}+\mathbf{P}^*+\mathbf{\Pi})=\\
    &(-1)\times{\rm CR\ flux\ source\ term}
    \end{aligned}
\end{equation}

\begin{equation}
\begin{aligned}\label{energy_cons}
    \frac{\partial \mathcal{E}_{\rm gas}}{\partial t}+&\nabla\cdot\left[(\mathcal{E}_{\rm gas}+P^*)\vec{u}-\vec{B}(\vec{B}\cdot\vec{u})+\mathbf{\Pi}\cdot\vec{u}\right]=\\
    &(-1)\times{\rm CR\ energy\ source\ term}
    \end{aligned}
\end{equation}

\begin{equation}\label{induction}
    \frac{\partial \vec{B}}{\partial t}-\nabla\times(\vec{u}\times\vec{B})=0
\end{equation}

\begin{equation}\label{eq:Ecr_code}
\begin{aligned}
    &\frac{\partial \mathcal{E}_{\rm cr}}{\partial t}+\nabla\cdot\vec{F}_{\rm cr}={\rm CR\ energy\ source\ term}=\\
    &-\sigma_+\left[\vec{F}_{\rm cr}-(\mathcal{E}_{\rm cr}+P_{\rm cr})(\vec{u}+\vec{v}_A)\right]\cdot(\vec{u}_{\parallel}+\vec{v}_A)\\
    &-\sigma_-\left[\vec{F}_{\rm cr}-(\mathcal{E}_{\rm cr}+P_{\rm cr})(\vec{u}-\vec{v}_A)\right]\cdot(\vec{u}_{\parallel}-\vec{v}_A)\\
    &+\left[(\sigma_+^L-\sigma_+^R)(u_\parallel+v_A)+(\sigma_-^L-\sigma_-^R)(u_\parallel-v_A)\right]c\Delta P_{\rm cr}\\
    &+A\sigma_{\Omega}\left(\vec{F}_{\rm cr}\times\vec{b}\right)\cdot\vec{u}
    \end{aligned}
\end{equation}

\begin{equation}\label{eq:Fcr_code}
\begin{aligned}
    &\frac{1}{V_m^2}\frac{\partial \vec{F}_{\rm cr}}{\partial t}+\nabla\cdot\mathbf{P}_{\rm cr}={\rm CR\ flux\ source\ term}\\
    =&-\sigma_+\left[\vec{F}_{\rm cr}-(\mathcal{E}_{\rm cr}+P_{\rm cr})(\vec{u}+\vec{v}_A)\right]\cdot\vec{b}\vec{b}\\
    &-\sigma_-\left[\vec{F}_{\rm cr}-(\mathcal{E}_{\rm cr}+P_{\rm cr})(\vec{u}-\vec{v}_A)\right]\cdot\vec{b}\vec{b}\\
    &+c(\sigma_+^L-\sigma_+^R+\sigma_-^L-\sigma_-^R)\Delta P_{\rm cr}\vec{b}\\
    &+A\sigma_{\Omega}\left[\vec{F}_{\rm cr}-(\mathcal{E}_{\rm cr}+P_{\rm cr})\vec{u}\right]\times\vec{b}
    \end{aligned}
\end{equation}

\autoref{mass_cons}-\ref{induction} describe the usual MHD equations, with $\rho$, $\vec{u}$ and $\mathcal{E}_{\rm gas}$ respectively denoting gas density, velocity and total energy $\mathcal{E}_{\rm gas}\equiv\rho u^2/2+P_{\rm therm}/(\gamma-1)+B^2/2$. $\mathbf{P}^*=P^* \mathbf{I}$ is the total pressure tensor with $P^*\equiv P_{\rm therm}+B^2/2$ being the corresponding scalar. $\mathbf{\Pi}$ is the viscous stress tensor. This MHD part is operated in the same way as in the original \textit{Athena}++, except that we add the negative of the CR source terms to the gas momentum and energy equations at the end of each step.

\autoref{eq:Ecr_code} and \autoref{eq:Fcr_code} derived from the last section form our CR module, where $\mathcal{E}_{\rm cr}$, $\vec{F}_{\rm cr}$ and $\mathbf{P}_{\rm cr}$ are lab-frame CR energy density, energy flux and pressure tensor. $\sigma_\pm^{L,R}$ are scattering coefficients due to Alfv\'en waves, $\sigma_{\Omega}$ is a shorthand for $\Omega_{\rm cr}/c^2$, which is physically a large parameter, $\Delta P_{\rm cr}$ is the CR pressure anisotropy to be supplied by the user as a function of environmental parameters, $A$ is a prefactor that accounts for the average gamma factor $\left<\gamma\right>$ in \autoref{Lorentz} (for a 1GeV proton, $\gamma=1.06$ and thus $A\lesssim 1$), and $V_m$ is the reduced speed of light, which should not alter the simulation result as long as $V_m$ is significantly larger than any other signal speed \citep{Skinner2013,JiangOh2018,Hopkins2022}. For most simulations presented in this paper, we adopt $V_m=100$ while other characteristic speeds remain of order unity. In all cases, we ensure that $V_m$ is much greater than any other characteristic speed.

\subsection{HLLE Riemann Solver}
We use the HLLE Riemann solver to solve the transport terms of \autoref{eq:Ecr_code} and \autoref{eq:Fcr_code} (LHS) as a hyperbolic equation:

\begin{equation}\label{transport_LHS}
    \frac{\partial \vec{q}}{\partial t}+\nabla\cdot\mathbf{f}=0
\end{equation}
where the conserved variables and fluxes are:

\begin{equation}\label{q}
    \vec{q}=\begin{pmatrix}
        \mathcal{E}_{\rm cr}\\
        {F}_{\rm cr,x}/V_m\\
        {F}_{\rm cr,y}/V_m\\
        {F}_{\rm cr,z}/V_m\\
    \end{pmatrix},\ \mathbf{f}=\begin{pmatrix}
        {F}_{\rm cr,x} & {F}_{\rm cr,y} & {F}_{\rm cr,z}\\
        V_m P_{\rm cr,xx} & V_m P_{\rm cr,xy} & V_m P_{\rm cr,xz}\\
        V_m P_{\rm cr,yx} & V_m P_{\rm cr,yy} & V_m P_{\rm cr,yz}\\        
        V_m P_{\rm cr,zx} & V_m P_{\rm cr,zy} & V_m P_{\rm cr,zz}\\
    \end{pmatrix}
\end{equation}
where the pressure tensor takes the form $P_{{\rm cr},ij}=(\mathcal{E}_{\rm cr}-\Delta P_{\rm cr})\delta_{ij}/3+\Delta P_{\rm cr}b_i b_j$ and $b_i$ is the projection of $\vec{b}$ onto lab-frame coordinate axes. $\delta_{ij}$ is the usual Kronecker symbol.



To update the conserved variables in a cell $(k,j,i)$ by a finite-volume scheme, we first reconstruct the left $(\vec{q}_L)$ and right states $(\vec{q}_R)$ using piecewise-linear (second-order) reconstruction \citep{Stone08}. The same procedure is also applied to $\Delta P_{\rm cr}$ and magnetic fields, which enter the HLLE fluxes through $P_{{\rm cr},ij}$. Using these reconstructed quantities, we then compute the HLLE fluxes at the cell interfaces to obtain their divergence. For example, the HLLE flux at the $(k,j,i-1/2)$ interface is:

\begin{equation}\label{eq:HLLEflux}
\begin{aligned}
    \mathbf{F}_{i-1/2}^{\rm HLLE}&=\frac{V^+ \mathbf{f}_x(\vec{q}_{L,i-1/2})-V^-\mathbf{f}_x(\vec{q}_{R,i-1/2})}{V^+-V^-}\\
    &+\frac{V^+V^-}{V^+-V^-}\left(\vec{q}_i-\vec{q}_{i-1}\right)
    \end{aligned}
\end{equation}
where $\mathbf{f}_x$ is the $x$ component (first column) of the flux tensor $\mathbf{f}$ calculated by substituting the matrix elements with corresponding left and right states, and $V^+$ and $V^-$ are the maximum and minimum wave speeds employed in the HLLE solver. Taking the transport terms split away from the source terms, the eigenvalues of hyperbolic \autoref{transport_LHS} give signal speeds $V^\pm=\pm V_m/\sqrt{3}$. However, as pointed out in \cite{Audit2002,Jiang2013}, such large signal speed can lead to excessive numerical diffusion, particularly in the regime dominated by scattering opacity. Inspired by numerical studies for RHD, \cite{JiangOh2018} introduced a smooth transition between the free-streaming wave speed $V_m/\sqrt{3}$ and the diffusion-limit speed $V_m/\tau$ (where $\tau$ is the optical depth) to reduce the signal speed thereby numerical diffusion. 

To address the issue of excessive numerical diffusion in a way that is directly motivated by the system at hand, we have analyzed typical waves in the CR-MHD system, and apply the results to set the maximum wave speed $V^+$ in our HLLE solver ($V^-$ is set to be $-V^+$).

More specifically, we adopt the following
\begin{equation}
    \begin{aligned}
        &V^+=\\
        &\begin{cases}
      V_m/\sqrt{3}, & \sigma\Delta x/2\pi<\sqrt{3}/{V_m} \\
      2\times {\rm max}(C_s, v_A), & \sqrt{3}/{V_m}<\sigma\Delta x/2\pi<{1}/10{C_{\rm max}}\\
      2\times C_{\rm max}, &{1}/10{C_{\rm max}}<\sigma\Delta x/2\pi
      
    \end{cases}
    \end{aligned}
\end{equation}
where $\sigma$ stands for the summation of all the four scattering coefficients here, $\Delta x$ is the cell length in the direction being calculated, $C_{\rm max}\equiv\sqrt{C_s^2 +C_{\rm cr}^2 +v_A^2}$ with $C_s^2\equiv \gamma_{\rm gas}P_{\rm therm}/\rho$ and $C_{\rm cr}^2\equiv (4/3)P_{\rm cr}/\rho=\gamma_{\rm cr}P_{\rm cr}/\rho$ respectively being the gas sound speed and CR sound speed. The factor 2 in the last two cases serves as a safety factor. The detailed derivation explaining the transition in wave speeds for different regimes and the reasoning for the safety factor can be found in \autoref{append:waves}.

\subsection{Source Terms}
On the RHS of our CR equations we have two types of source terms. One is only associated  with $\Delta P_{\rm cr}$, which we prescribe, but the other contains CR variables $\mathcal{E}_{\rm cr}$ and $\vec{F}_{\rm cr}$ being solved. We directly add the first type as explicit source terms after updating the advection terms, while we treat the second type implicitly to avoid stiffness.

Although the implementation of explicit source terms is straightforward, we caution that it must be properly combined with the implicit integrators so that the overall algorithm can become more accurate than a first-order method \citep{Huang2022}. We denote the conserved CR variables after updating explicit source terms as $\mathcal{E}_{\rm cr}^*$, $\vec{F}_{\rm cr}^*$ and $\vec{q}^*$, which includes contributions from both advection and explicit source terms. We combine them with the implicit integrators described below.

As scattering is primarily along magnetic field lines for the implicit part, in each cell we first rotate the coordinate system by a rotation matrix $\mathbf{R}$ as in \cite{JiangOh2018}:
\begin{equation}
    \begin{aligned}
        \mathbf{R}=\begin{pmatrix}
            {\rm cos}\phi_B {\rm sin}\theta_B & {\rm sin}\phi_B{\rm sin}\theta_B & {\rm cos}\theta_B \\
            -{\rm sin}\phi_B & {\rm cos}\phi_B & 0\\
            -{\rm cos}\phi_B{\rm cos}\theta_B & -{\rm sin}\phi_B{\rm cos}\theta_B & {\rm sin}\theta_B
        \end{pmatrix}
    \end{aligned}
\end{equation}
with
\begin{equation}
\begin{aligned}
    &{\rm cos}\phi_B\equiv\frac{B_x}{\sqrt{B_x^2+B_y^2}},\ {\rm sin}\phi_B\equiv\frac{B_y}{\sqrt{B_x^2+B_y^2}}\\
    &{\rm cos}\theta_B \equiv B_z/B,\ {\rm sin}\theta_B\equiv \sqrt{1-{\rm cos}^2\theta_B}
    \end{aligned}
\end{equation}
so that $\hat{x}'\parallel\vec{b}$ in the new coordinate system (prime $'$ means axial components measured in the new coordinate system).

We then reassemble the scattering coefficients and the Lorentz factor into scattering tensors $\overleftrightarrow{\sigma}_\pm$, whose matrix forms in the new coordinate system are:

\begin{equation}\label{sigma_matrix}
\overleftrightarrow{\sigma}_\pm\equiv\begin{pmatrix}
       \sigma_\pm & 0 & 0\\
       0 & 0 & -A\sigma_{\Omega}/2\\
       0 & A\sigma_{\Omega}/2 & 0
    \end{pmatrix}
\end{equation}
Tensor-form scattering coefficients are also convenient for future extensions such as adding perpendicular scattering terms due to field line wandering. Note that it is only in the coordinate system with $\hat{x}'\parallel\vec{b}$ that $\overleftrightarrow{\sigma}_\pm$ have such a straightforward matrix form.

With these prerequisites, below we introduce our implicit schemes where the design of algorithm directly follows the dust-gas drag module in \cite{Huang2022}.

\subsubsection{Backward Euler method}

We first introduce the simple first-order implicit integrator to be combined with the Runge-Kutta (RK) 1 time integrator in \textit{Athena}++:
\begin{equation}
\begin{aligned}
        &\frac{\mathcal{E}_{\rm cr}^{(n+1)}-\mathcal{E}_{\rm cr}^{(n)}}{\Delta t}=\frac{\mathcal{E}_{\rm cr}^{*(n)}-\mathcal{E}_{\rm cr}^{(n)}}{\Delta t}\\
        &-{\vec{w}_{+,\parallel}^{(n)\ T}}\cdot\overleftrightarrow{\sigma}_+^{(n)}\cdot\left[\vec{F}_{\rm cr}^{(n+1)}-\frac{4}{3}\mathcal{E}_{\rm cr}^{(n+1)}\vec{w}_+^{(n)}\right]\\
        &-\vec{w}_{-,\parallel}^{(n)\ T}\cdot\overleftrightarrow{\sigma}_-^{(n)}\cdot\left[\vec{F}_{\rm cr}^{(n+1)}-\frac{4}{3}\mathcal{E}_{\rm cr}^{(n+1)}\vec{w}_-^{(n)}\right]
    \end{aligned}
\end{equation}
\\

\begin{equation}
    \begin{aligned}
    &\frac{1}{V_m^2}\frac{\vec{F}_{\rm cr}^{(n+1)}-\vec{F}_{\rm cr}^{(n)}}{\Delta t}=\frac{\vec{F}_{\rm cr}^{*(n)}-\vec{F}_{\rm cr}^{(n)}}{V_m^2\Delta t}\\
    &-\overleftrightarrow{\sigma}_+^{(n)}\cdot\left({\vec{F}_{\rm cr}^{(n+1)}-\frac{4}{3}\mathcal{E}_{\rm cr}^{(n+1)}\vec{w}_+^{(n)}}\right)\\
    &-\overleftrightarrow{\sigma}_-^{(n)}\cdot\left({\vec{F}_{\rm cr}^{(n+1)}-\frac{4}{3}\mathcal{E}_{\rm cr}^{(n+1)}\vec{w}_-^{(n)}}\right)
    \end{aligned}
\end{equation}
where superscripts denote steps. $\vec{w}_\pm^{n}\equiv\vec{u}^{n}\pm\vec{v}_A^{n}$ are Alfv\'en wave velocities as before, and the $\parallel$ subscript means components parallel with magnetic fields.

We can also rewrite the implicit source terms into a matrix form $\mathbf{S}\vec{q}$
where $\vec{q}$ is the CR variable vector as defined in \autoref{q} and $\mathbf{S}$ is the matrix whose elements in the $x'\parallel \vec{b}$ coordinate are 

\begin{equation}
    \begin{aligned}
    &\mathbf{S}=\\
    &\begin{pmatrix}
    \frac{4}{3}\sum_{i=\pm}\sigma_i w_i^2 & -\sum_{i=\pm}\sigma_i w_i & -A\sigma_{\Omega}u_z' & -A\sigma_{\Omega}u_y'\\
    \frac{4}{3}\sum_{i=\pm}\sigma_i w_i & -V_m\sum_{i=\pm}\sigma_i & 0 & 0\\
    -\frac{4}{3}A\sigma_{\Omega}V_m u_z' & 0 & 0 & A\sigma_{\Omega}V_m\\
    \frac{4}{3}A\sigma_{\Omega}V_m u_y' & 0 & -A\sigma_{\Omega}V_m & 0
    \end{pmatrix}
    \end{aligned}
\end{equation}
so that the backward Euler method can be expressed as:

\begin{equation}
    \vec{q}^{(n+1)}=\left(\mathbf{I}-\Delta t\mathbf{S}^{(n)}\right)^{-1}\left(\vec{q}^{(n)}+\Delta \vec{q}_{\rm af,src}^{(n)}\right)
\end{equation}
where $\mathbf{I}$ is the unit matrix and $\Delta\vec{q}_{\rm af,src}\equiv \vec{q}^*-\vec{q}^{(n)}$. Both scattering coefficients and MHD quantities are treated as constant parameters here. They are calculated at the beginning of each step.

\subsubsection{Implicit van Leer 2 (VL2) method}
We also construct a second-order implicit VL2 integrator, used in tandem with the two-stage VL2 and RK2 integrators in \textit{Athena}++. Using the notation above, our implicit VL2 method can be described as below:

{\bf Stage 1}: evolve the system for half a time step $\Delta t/2$ in the same way as the backward Euler method above:
\begin{equation}
    \vec{q}^{(n+1/2)}=\left(\mathbf{I}-\frac{\Delta t}{2}\mathbf{S}^{(n)}\right)^{-1}\left(\vec{q}^{(n)}+\Delta \vec{q}_{\rm af,src}^{(n)}\right)
\end{equation}
then update the conserved variables of the MHD gas accordingly.

{\bf Stage 2}: use the updated quantities at the $n+1/2$ step to evolve the system from step $n$ to $n+1$. Here we directly list the eventual scheme while the detailed derivation can be found in the Appendix B\footnote{The derivation in \cite{Huang2022} was based on a general form of source terms, while our situation is a special case where the implicit source terms linearly depend on the conservative variables.} of \cite{Huang2022}:

\begin{equation}
    \begin{aligned}
        &\vec{q}^{(n+1)}-\vec{q}^{(n)}=\\
        &\mathbf{\Lambda}^{-1}\left(\mathbf{I}-\frac{\Delta t}{2}\mathbf{S}^{(n+1/2)}\right)\Delta t \left(\mathbf{S}^{(n+1/2)}\vec{q}^{(n)}+\Delta\vec{q}_{\rm af,src}^{(n+1/2)}\right)
    \end{aligned}
\end{equation}
with $\mathbf{\Lambda}$ being:

\begin{equation}
    \mathbf{\Lambda}=\mathbf{I}-\left(\mathbf{I}-\frac{\Delta t}{2}\mathbf{S}^{(n+1/2)}\right)\Delta t \mathbf{S}^{(n+1/2)}
\end{equation}

In practice, we find that the second-order method offers much better numerical stability and robustness. We thus recommend its use. For most tests presented in this paper, we employ the VL2 integrator.

\subsection{Stability Criterion for Time Step}
It is well known that for equations describing a quantity streaming down its gradient, grid-scale oscillations can rapidly grow and overwhelm the true solution. We indeed observe this behavior in our 1D streaming tests (see more details in \autoref{subsec:1Dstreaming}), where the scattering coefficients take the form given in \autoref{eq:stscatt}. As $\mathcal{E}_{\rm cr}$ streams down from a maxima, $\nabla P_{\rm cr}$ can flip sign due to an overshoot, which also flips the sign of $F_{\rm cr}$, then making the sign of $\nabla P_{\rm cr}$ change even more frequently, generating numerical oscillations. The magnitude of such oscillations may be amplified or suppressed as simulation evolves, depending on the chosen time step.  

To overcome this serious issue, \cite{Sharma10} regularized the one-moment CR transport equation by adding a numerical diffusion term. A disadvantage of this method is that it introduces artificial diffusion that can be hard to distinguish from the physical one, and also increases the computational expenses. \cite{JiangOh2018} instead removed this sign-flipping term from the RHS by combining it with the advection term on the LHS, and identified the modified source term as the effective scattering in the co-moving frame. This treatment is numerically convenient, but their formulation implicitly assumes that the scattering mechanism is solely subject to the CRSI and there is only one direction for Alfv\'en waves to propagate. Therefore, their method is inherently difficult to extend to the more general form we aim to consider in this work.

While our formulation provides a more general framework that can incorporate diverse physics, it is accompanied by the sign-flipping terms whose numerical influence must be examined. To this end, we perform a traditional von Neumann analysis for our two-moment equations, obtaining  guidelines to set the time step such that the system is numerically stable. For scattering coefficients in the form of \autoref{eq:stscatt}, the stability criterion is
\begin{equation}\label{eq:stabcriterion}
    \Delta t<\frac{\sqrt{3}}{1+4\widetilde{\sigma}_{0,\rm st}}\frac{\Delta x}{V_m}.
\end{equation}
A detailed derivation can be found in \autoref{append:timestep}.

When $\widetilde{\sigma}_{0,\rm st}\lesssim 1$, the $\Delta t$ given here is comparable to the one limited by the transport term $(\Delta t< \sqrt{3}\Delta x/V_m)$, and therefore does not impose a significant computational cost for resolving CR transport. This regime is typical when CR scattering is mediated by CRSI-driven waves damped by ion–neutral collisions in ISM, for which equating \autoref{eq:sigma_CRSI} and \autoref{eq:stscatt} yields $\widetilde{\sigma}_{0,\rm st}^{\rm IN} \sim\Omega_{\rm cr}(\rho _{\rm cr}/\rho_{\rm ion})/\nu_{\rm damp}\lesssim 10^{-2}$ \citep[see e.g.][for representative ISM parameters]{Armillotta2022}.


For non-linear Landau damping acting efficiently in ionized gas, the magnitude of the corresponding scattering coefficient can exceed $\sigma_{\rm IN}$ by 3-4 orders of magnitude \citep[for reference, see equation 39 and 40 of][]{Hix2025}. Accordingly, the equivalent dimensionless parameter $\widetilde{\sigma}_{0,\rm st}^{\rm NLL}$ satisfies $\widetilde{\sigma}_{0,\rm st}^{\rm NLL}/\widetilde{\sigma}_{0,\rm st}^{\rm IN}=\sigma_{\rm NLL}/\sigma_{\rm IN}\sim 10^{3-4}$, leading to a more restrictive $\Delta t$. Nevertheless, the resulting $\Delta t$ remains computationally manageable in many practical cases.

We caution that this criterion was derived for scattering coefficients of the form given in \autoref{eq:stscatt} (CRSI driving balanced by ion–neutral damping). Although one can formally compute an equivalent $\widetilde{\sigma}_{0,\rm st}$ for other damping mechanisms, such as the non-linear Landau damping discussed above, there is no guarantee that the resulting $\Delta t$ will rigorously ensure numerical stability. In practice, however, our test indicates that the criterion remains approximately applicable (see \autoref{subsec:1Dnlld}).

We adopt this time step criterion in all simulations with streaming-dominated scattering coefficients (except for \autoref{fig:1Dstability} and \autoref{fig:1Dstability_vl2} where the stability condition is explicitly tested).

\section{Numerical Tests of the CR subsystem}\label{sec:4}
We demonstrate the performance of our CR-MHD module in \textit{Athena}++ via a series of numerical tests.
Similar to \cite{JiangOh2018}, we first test the CR subsystem in this section where the CR feedback to MHD gas is turned off. Additionally, except in \autoref{subsec:CRPAItest} we neglect the $\mathcal{E}_{\rm cr}$ source term so that the CR energy is conserved and the subsystem is easier to analyze. In other words, for  \autoref{subsec:1Dstreaming}-\autoref{subsec:sphCRsttest}, $\mathcal{E}_{\rm cr}$ is purely transported by the CR energy flux $\vec{F}_{\rm cr}$:
\begin{equation}\label{eq:consEcr}
    \frac{\partial \mathcal{E}_{\rm cr}}{\partial t}+\nabla\cdot\vec{F}_{\rm cr}=0
\end{equation}
simply for test purposes.

\subsection{1D CR Streaming (linear damping)}\label{subsec:1Dstreaming}
One crucial physical element which must be captured is that a non-zero $\nabla P_{\rm cr}$ can drive CR bulk motion down the gradient, with the streaming speed limited by Alfv\'en waves driven by CRSI. Following \autoref{eq:sigma_CRSI}, we set the scattering coefficient to 
\begin{equation}\label{eq:stscatt}
\sigma_{\pm}^{L,R}=\frac{\widetilde{\sigma}_{0,\rm st}}{2}\frac{|\nabla P_{\rm cr}|}{v_A P_{\rm cr}}\Theta(\mp\nabla P_{\rm cr}\cdot\vec{b}),
\end{equation}
where $\Theta$ is the Heaviside function.

Here the dimensionless prefactor $\widetilde{\sigma}_{0,\rm st}$ account for the factor $\sim\Omega_{\rm cr}(\rho _{\rm cr}/\rho_{\rm ion})/\nu_{\rm damp}$ in \autoref{eq:sigma_CRSI}. In reality, it can take on a range of values depending on the CR density and environmental parameters. For simplicity, here we treat $\widetilde{\sigma}_{0,\rm st}$ as a constant, encoding the overall amplitude on top of the key scaling $|\nabla P_{\rm cr}|/P_{\rm cr}$ that reflects the characteristics of streaming.

Under this prescription, the steady-state flux equation yields
\begin{equation}\label{eq:frontspeed}
F_{\rm cr}=-\frac{(1+4\widetilde{\sigma}_{0,\rm st})}{3\widetilde{\sigma}_{0,\rm st}} v_A \mathcal{E}_{\rm cr}{\rm sgn}\left(\frac{\partial \mathcal{E}_{\rm cr}}{\partial x}\right)+\frac{4}{3}u\mathcal{E}_{\rm cr}
\end{equation}
in regions with $|\nabla \mathcal{E}_{\rm cr}|$ far from $0$. We thus set a triangular profile for the initial CR energy density $\mathcal{E}_{\rm cr}(t=0,x)=2-|x|$ so that $|\nabla \mathcal{E}_{\rm cr}|$ is constant. We set $v_A=1$  with $\vec{b} = \hat{x}$, and the background MHD gas is static with all fluid variables fixed since the CR feedback to gas is turned off. Plugging the steady-state $F_{\rm cr}$ back to the conservative $\mathcal{E}_{\rm cr}$ equation, we obtain traveling-wave equation except at $\partial \mathcal{E}_{\rm cr}/\partial x$ discontinuities. Consequently, during the evolution of the system, $|\nabla \mathcal{E}_{\rm cr}|$ is either $1$ at the propagating fronts, or $0$ in the central flat part originating from the initial discontinuity. For the default test here we use $\widetilde{\sigma}_{0,\rm st}=1$, and set $\Delta t=0.3\Delta x/V_m$ according to our stability criterion (\autoref{eq:stabcriterion}).

\begin{figure}
\includegraphics[scale=0.475]{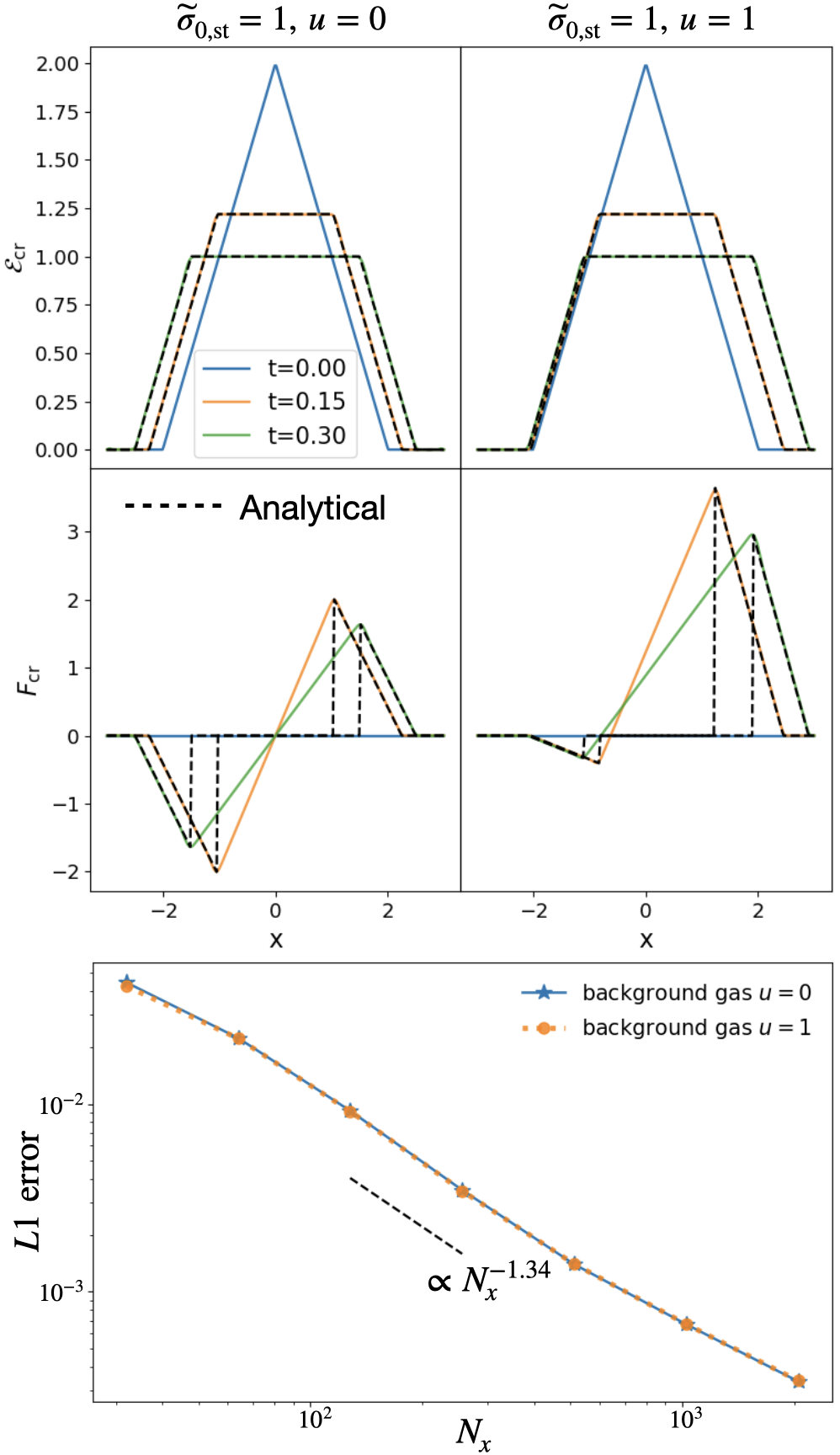}
    \caption{Numerical results of 1D streaming tests from \autoref{subsec:1Dstreaming}. {\bf Top left:} $\mathcal{E}_{\rm cr}$ and $F_{\rm cr}$ in a static background gas $(u=0)$. Colored lines denote snapshots at increasing simulation times, while dashed curves show analytical solutions, validating agreement between numerical and theoretical results. {\bf Top right:} Identical setup with a uniformly advected background gas $(u=1)$, introducing additional $F_{\rm cr}$ from gas motion. The $\mathcal{E}_{\rm cr}$
  profile drifts at $4u/3$, consistent with expectations. {\bf Bottom:} Spatial convergence of $L1$ errors $(\sum_{i=1}^{N_x}|\mathcal{E}_{\rm cr,i}-\mathcal{E}_{\rm cr,theory}|/N_x)$ at $t=0.3$. Errors roughly scale as $\propto N_x^{-1.34}$.}
    \label{fig:1Dstream}
\end{figure}

\cite{JiangOh2018} calculated the analytical solution of this problem, and by comparing it with our simulations in the left panel of \autoref{fig:1Dstream}, we see our numerical results match very well with the analytical solutions except for the central regions in the $F_{\rm cr}$ panel. In the central plateau where $\nabla \mathcal{E}_{\rm cr}=0$, the time derivative $\partial F_{\rm cr}/\partial t$ can no longer be neglected and the theoretical expectation above is not valid. To test that numerical diffusion produced by the advection term does not overwhelm true physical solutions, we do the same streaming test in a moving gas background, where the background gas uniformly moves at a constant velocity $u=1$. This background movement should only introduce an additional $4u\mathcal{E}_{\rm cr}/3$ term in the steady $F_{\rm cr}$ (\autoref{eq:frontspeed}), causing the center of $\mathcal{E}_{\rm cr}$ profile to drift at $4u/3$ on top of the previous solution. In the right panel of \autoref{fig:1Dstream} we show numerical results for the moving gas background, which agrees well with the analytic steady solution. We also measure the $L1$ errors in $\mathcal{E}_{\rm cr}$ at $t=0.3$ in our simulations with varying resolutions, and show in the bottom panel that the errors roughly decrease at a rate $\propto N_x^{-1.34}$ where $N_x$ is the cell number. Although we use a second-order scheme for this test, the scattering coefficients $\sigma_\pm^{L,R}$ depend on $\mathcal{E}_{\rm cr}$ but are treated as fixed parameters here. As a result, the method is not fully implicit, and the actual spatial convergence lies between first and second order.

\begin{figure*}
    \centering
    \includegraphics[scale=0.405]{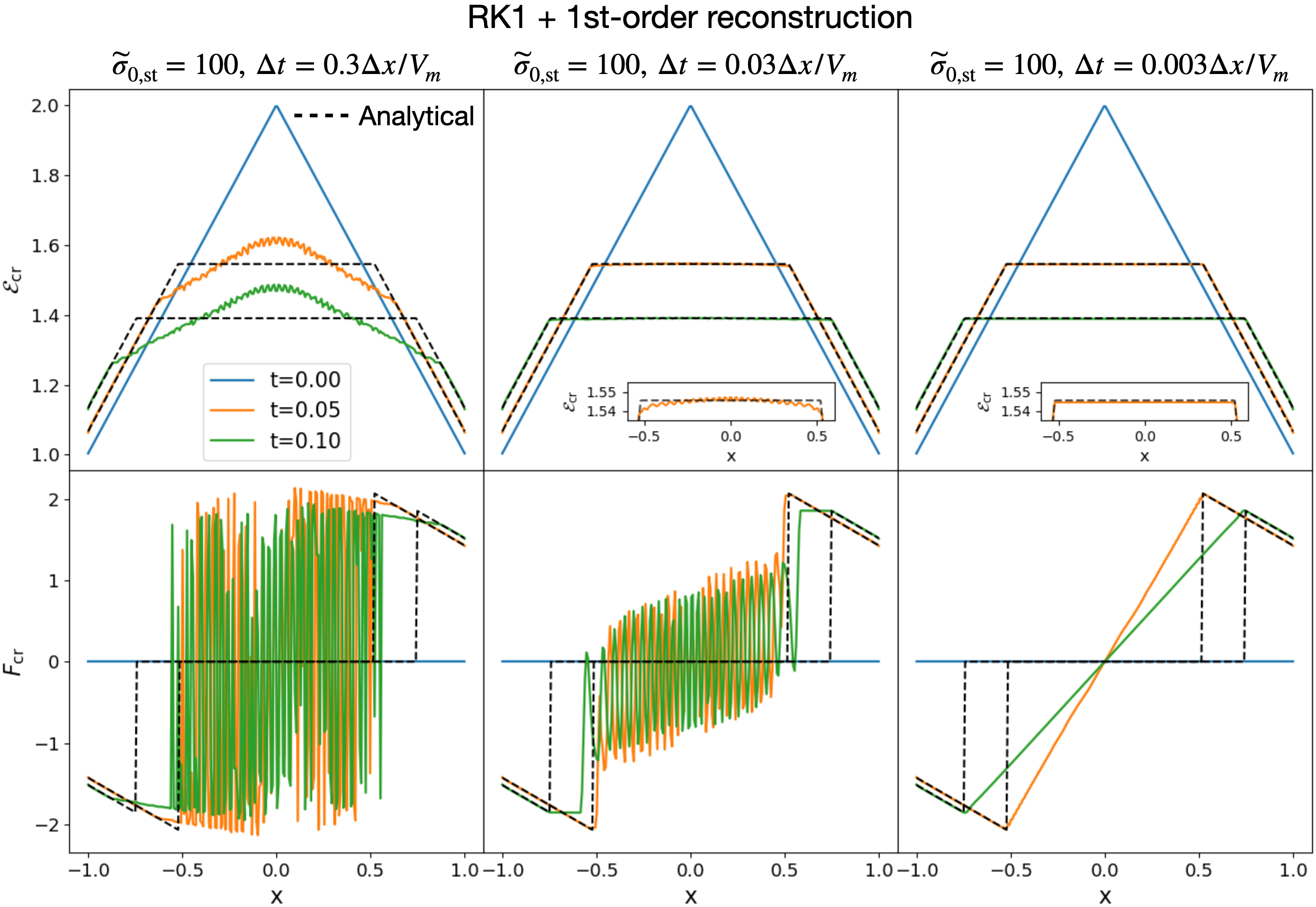}
    \caption{Numerical results of the 1D streaming test using the same setup as in \autoref{fig:1Dstream}, except for elevated scattering coefficients $\widetilde{\sigma}_{0,\rm st}$ to illustrate the performance of our stability criterion. Time steps $\Delta t$ progressively decrease from left to right, as indicated by the titles. We see numerical instability and deviation from analytical solutions persist until the stability condition $\Delta t<\sqrt{3}\Delta x/(1+4\widetilde{\sigma}_{0,\rm st})V_m$ is satisfied in the rightmost column.}
    \label{fig:1Dstability}
\end{figure*}

Numerical results remain stable in our tests with $\widetilde{\sigma}_{0,\rm st}=1$. However, as $\widetilde{\sigma}_{0,\rm st}$ increases in \autoref{eq:stscatt}, the magnitude of overshoot around $\nabla P_{\rm cr}$ extrema also grows, leading to stronger numerical oscillations. This behavior demands a tighter time step constraint to gradually suppress the oscillations rather than let them amplify into instabilities, consistent with the stability condition given by \autoref{eq:stabcriterion}. To further test this criterion, we increase the dimensionless prefactor $\widetilde{\sigma}_{0,\rm st}$ by a factor of 100, such that the original timestep $\Delta t$ now violates the stability condition. In this experiment (left column of \autoref{fig:1Dstability}), we see both $\mathcal{E}_{\rm cr}$ and $F_{\rm cr}$ are choppy in the central plateau instead of connecting the left and right fronts smoothly. We then reduce $\Delta t$ by a factor of 0.1, as shown in the middle panels, which still can not fully stabilize the system. With a further reduction of $\Delta t$ by a factor 0.1, so that the stability condition is satisfied, the right column shows that stability is again recovered. This result lends support to our stability criterion as a useful time-step constraint for mitigating numerical instabilities in scenarios involving streaming-driven scattering terms.

Note that the simulations demonstrated in \autoref{fig:1Dstability} were performed with a 1st-order method, so that the numerical scheme aligns with our von Neumann stability analysis presented in \autoref{append:timestep}. All other simulations in this paper were run with a 2nd-order scheme (VL2 integrator $+$ piecewise linear reconstruction). In practice we find that second-order methods usually deliver better performance with relaxed time-step constraints (see \autoref{fig:1Dstability_vl2}). Nonetheless, we still prescribe our stability condition as a practical guideline.

\subsection{1D CR Streaming (non-linear Landau damping)}\label{subsec:1Dnlld}


CR transport is also sensitive to the wave damping process, which balances wave growth, modulates CR scattering, and ultimately determines the degree of coupling between CRs and the MHD gas \citep{Armillotta2022,Thomas2025b}. In our formulation, the influence of various damping mechanisms is essentially encapsulated into scattering coefficients. When CR scattering is mediated by MHD waves driven by CRSI but damped primarily through nonlinear Landau damping, the scattering coefficients take the form \citep[e.g.,][]{Armillotta2022,Hix2025}:
\begin{equation}\label{eq:sigma_nll}
\sigma_{\pm}^{L,R}=\frac{\widetilde{\chi}_{0,\rm NLL}}{2} P_{\rm therm}^{-1/4}\left(\frac{e\left|\nabla P_{\rm cr}\right|}{mc^4}\right)^{1/2}\Theta(\mp\nabla P_{\rm cr}\cdot\vec{b})
\end{equation}
where $e/m$ is the charge-to-mass ratio of a CR particle and $\widetilde{\chi}_{0,\rm NLL}$ is a dimensionless constant prefactor absorbing microphysical uncertainties or calibration.

In this subsection, we test the nonlinear Landau damping mechanism by performing a 1D test similar to that in \autoref{subsec:1Dstreaming}, again holding the MHD variables fixed. Since $P_{\rm therm}=1$ is a constant, we group the associated environmental factors into a new dimensional constant $\chi_{0,\rm NLL}\equiv\widetilde{\chi}_{0,\rm NLL}P_{\rm therm}^{-1/4}(e/m)^{1/2}c^{-2}$ for numerical convenience. The resulting scattering coefficients are then

\begin{equation}\label{eq:sigma_nll_code}
\sigma_{\pm}=\chi_{0,\rm NLL}\sqrt{\left|\nabla P_{\rm cr}\right|}\Theta(\mp\nabla P_{\rm cr}\cdot\vec{b}).
\end{equation}

The steady-state $F_{\rm cr}$ in this case (taking the gas velocity to be zero) is solved to be:
\begin{equation}
F_{\rm cr}=-\left(\frac{\sqrt{|\nabla P_{\rm cr}|}}{\chi_{0,\rm NLL}}+\frac{4}{3}v_A \mathcal{E}_{\rm cr}\right){\rm sgn}(\nabla P_{\rm cr}\cdot\vec{b})
\end{equation}

We initialize the $\mathcal{E}_{\rm cr}$ profile as:
\begin{equation}
    \mathcal{E}_{\rm cr}(0,x)=\frac{1}{\mathcal{E}_0^{-1}+3\chi_{0,\rm NLL}^2 v_{0}^2|x|}
\end{equation}
where $\mathcal{E}_0$ and $v_0$ are constant parameters and $x\in (-L_{\rm box}/2,L_{\rm box}/2)$. One can find initially $\sqrt{|\nabla P_{\rm cr}|}=\chi_{0,\rm NLL}v_{0} \mathcal{E}_{\rm cr}$, such that $F_{\rm cr}=-\left(v_0+4v_A/3\right)\mathcal{E}_{\rm cr}{\rm sgn}(\nabla P_{\rm cr}\cdot\vec{b})$. Assuming $v_A$ and $v_0$ to be constant, this form of $F_{\rm cr}$ makes traveling waves the solutions to the $\mathcal{E}_{\rm cr}$ equation, and the fronts propagate at the speeds of $\pm(v_0 +4v_A/3)$.

As the initial $\mathcal{E}_{\rm cr}$ peak streams outwards, a plateau is developed in the central region of the domain, where our estimation of the steady-state $F_{\rm cr}$ fails because of the zero $\nabla P_{\rm cr}$. Denoting the plateau region as $(-x_p, x_p)$, the time evolution of the $\mathcal{E}_{\rm cr}$ profile is thus:

\begin{equation}\label{eq:Ecr_nlldamping}
    \mathcal{E}_{\rm cr}(t,x)=    \begin{cases}
      \mathcal{E}_{\rm cr}\left[0,|x|-\left(v_0+\frac{4}{3}v_A\right)t\right], & |x|\geq x_p\\
      \mathcal{E}_{\rm cr}(t,x_p), & |x|< x_p
    \end{cases}
\end{equation}

The value of $x_p$ can be calculated by $\mathcal{E}_{\rm cr}$ conservation of the system:

\begin{equation}
\begin{aligned}
    &\int_0^t F_{\rm cr}(t,L_{\rm box}/2)dt\\
    &+ \int_{x_p}^{L_{\rm box}/2}\mathcal{E}_{\rm cr}(t,x)dx+\mathcal{E}_{\rm cr}(t,x_p)x_p\\
    &=\int_0^{L_{\rm box}/2}\mathcal{E}_{\rm cr}(0,x)dx
    \end{aligned}
\end{equation}

In this equation, the first term represents the total CR energy flowing through the outflow boundaries, while the second and third terms separately represent the total CR energy in the outskirt and the central plateau. Their sum should equal the total initial CR energy since $\mathcal{E}_{\rm cr}$ is conserved in these tests. The range of the central plateau $x_p$ can be thus obtained from the resulting transcendental equation:

\begin{equation}
\begin{aligned}
\frac{x_p}{(3\chi_{0,\rm NLL}^2v_0^2\mathcal{E}_0)^{-1}+x_p-(v_0+4v_A/3)t}\\
    = {\rm log}\left[1+\frac{x_p-(v_0+4v_A/3)t}{(3\chi_{0,\rm NLL}^2v_0^2\mathcal{E}_0)^{-1}}\right]
    \end{aligned}
\end{equation}

\begin{figure}
\includegraphics[scale=0.48]{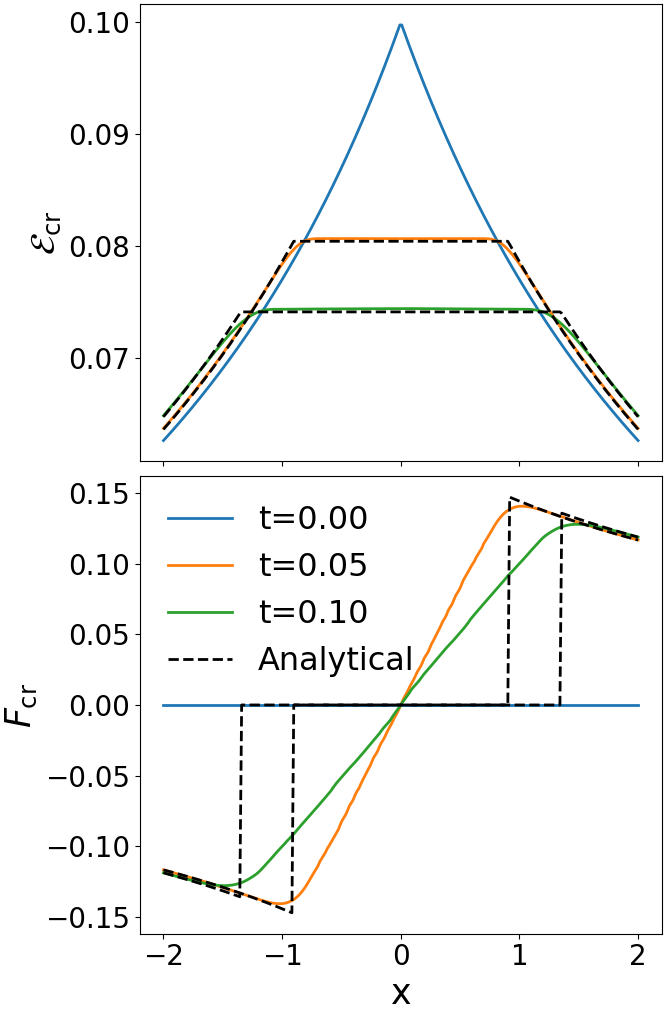}
    \caption{Numerical results of 1D streaming tests from \autoref{subsec:1Dnlld}. This is similar to \autoref{fig:1Dstream} but the damping mechanism follows the form expected for non-linear Landau damping, with the scattering coefficients defined in  \autoref{eq:sigma_nll}. Dashed lines denote our theoretical predictions, which align well with the simulation data at the propagating fronts. Deviations occur in the central plateau, where $\nabla P_{\rm cr}\rightarrow 0$ violates our theoretical assumption of a non-negligible $\nabla P_{\rm cr}$. As a result $\mathcal{E}_{\rm cr}$ and $F_{\rm cr}$ smoothly transition to the propagating fronts.}
    \label{fig:1Dstreaming_nll}
\end{figure}

In our test problem, we set $\chi_{0,\rm NLL}=2$, $v_0=0.5$, $\mathcal{E}_0=0.1$ and $L_{\rm box}=4$. The MHD gas properties are identical to those in \autoref{subsec:1Dstreaming}, so that $v_A=1$. We compare the reference solution \autoref{eq:Ecr_nlldamping} with numerical results in \autoref{fig:1Dstreaming_nll}, and see excellent agreement. Therefore, our code can incorporate various types of scattering coefficients which essentially correspond to different microphysical processes.

We comment that our stability criterion, although not designed for this test, in practice still works fine with a reduction factor of $\sim 2.35$. Directly equating \autoref{eq:stscatt} and \autoref{eq:sigma_nll}, the effective $\widetilde{\sigma}_{0,\rm st}^{\rm NLL}$ in this test is about 0.67. Substituting it directly back to the stability criterion \autoref{eq:stabcriterion} gives the stable time step $\Delta t=0.47\Delta x/V_m$, while we roughly achieve stability with $\Delta t=0.2\Delta x/V_m$ used in the test above. We used a spatial resolution $\Delta x=1/64$ here. However, we caution that, details of numerical requirements of different forms of scattering coefficients (particularly on the time step) await future explorations.

\subsection{1D CR Diffusion}\label{sec:diff_test}
Another important origin of CR scattering is MHD waves generated via a cascade from extrinsic (i.e. ambient ISM or CGM) turbulence, which has no preferred direction of propagation. The resulting CR transport thus becomes a diffusion-like process.

To test this transport regime, we set all the four scattering coefficients equal and constant: $\sigma_+^L=\sigma_+^R=\sigma_-^L=\sigma_-^R\equiv\sigma_{0,\rm diff}/4$ with $\sigma_{0,\rm diff}$ denoting their summation. By solving for the steady $F_{\rm cr}$ and plugging it back into the CR energy density \autoref{eq:consEcr}, we obtain

\begin{equation}
    \frac{\partial \mathcal{E}_{\rm cr}}{\partial t}+\frac{4}{3}u\frac{\partial \mathcal{E}_{\rm cr}}{\partial x}-\frac{1}{3\sigma_{0,\rm diff}}\frac{\partial ^2 \mathcal{E}_{\rm cr}}{\partial x^2}=0
\end{equation}
where gas velocity $u$ is assumed uniformly constant. Combining the first two terms as a comoving derivative $D\mathcal{E}_{\rm cr}/Dt$, we see it becomes a standard diffusion equation in the frame moving\footnote{The fact that $\mathcal{E}_{\rm cr}$ is advected at a different velocity from the background gas breaks Galilean invariance. The reason is that by neglecting the source term in \autoref{eq:consEcr} we ignored the work done by $P_{\rm cr}$ on the moving gas, and thus reach to a somewhat unphysical situation. After recovering the $\mathcal{E}_{\rm cr}$ source term, one can see in reality the $\mathcal{E}_{\rm cr}$ profile should flow at the same velocity as the background gas (provided the $v_A^2$ heating terms corresponding to the Fermi acceleration are omitted).} with the velocity $4u/3$. We initialize the system with a Gaussian profile $\mathcal{E}_{\rm cr}(0,x)={\rm exp}(-40x^2)$, and the theoretical solution to the diffusion equation can be obtained via Green's function method. When the background gas is static, \cite{JiangOh2018} wrote down the analytical result:

\begin{equation}
\begin{aligned}
    \mathcal{E}_{\rm cr}(t,x)=\frac{1}{\sqrt{1+160t/3\sigma_{0,\rm diff}}}{\rm exp}\left(\frac{-40x^2}{1+160t/3\sigma_{0,\rm diff}}\right)
    \end{aligned}
\end{equation}
and a translation from $x$ to $x-4ut/3$ extends the result to the case with a uniformly moving background gas.

\begin{figure}
\includegraphics[scale=0.5]{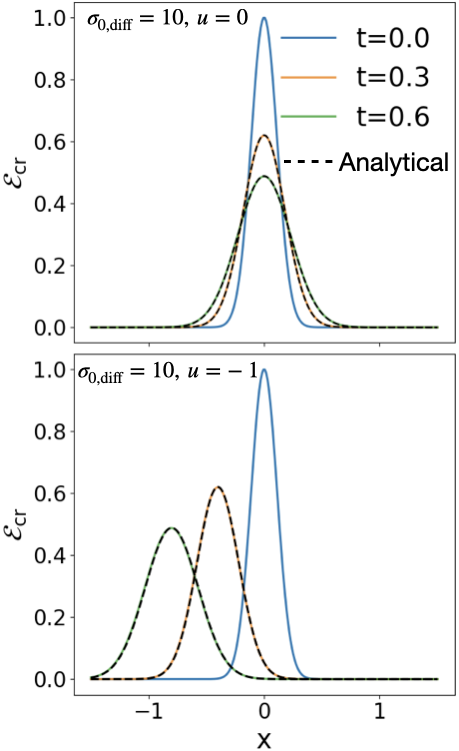}
    \caption{Numerical results of 1D diffusion tests described in \autoref{sec:diff_test}. {\bf Top panels:} Evolution of $\mathcal{E}_{\rm cr}$ profiles in a static MHD background. {\bf Bottom panels:} Evolution of $\mathcal{E}_{\rm cr}$ profiles in a uniformly moving MHD background (MHD velocity $u=-1$). }
    \label{fig:1Ddiff}
\end{figure}

In \autoref{fig:1Ddiff} we present the results of 1D diffusion tests, which closely align with theoretical expectations for both a static gas background and a uniformly moving background at a velocity $u=-1$. These results demonstrate that our code accurately handles this diffusive transport regime, with numerical diffusion remaining well below physical diffusion levels.

\subsection{Anisotropic CR Flux}\label{sec:aniso_flux}
CRs are
generally tied to field lines, with $F_{\rm cr}$ streaming along magnetic fields at macroscopic scales. This physical feature is reflected in the matrix form of the scattering tensor (\autoref{sigma_matrix}) where the two bottom right diagonal terms are 0. However, the direction of magnetic fields generally varies spatially, and the CR inertia may thus cause a perpendicular component of CR flux. We add the corresponding macroscopic Lorentz force (\autoref{Lorentz}) to counteract this inertia. In reality, the gyro-frequency is very large, which numerically keeps the CR flux aligned with the magnetic fields in the comoving frame.

In this subsection, we test the performance of the Lorentz term with a commonly employed problem setup \citep[e.g.,][]{Parrish05,Sharma11,Pakmor16,JiangOh2018}, where the magnetic field is circular in a 2D Cartesian $(x,y)\in(-1,1)\times(-1,1)$ domain:

\begin{equation}
    B_x(x,y)=\frac{-y}{\sqrt{x^2+y^2}},\ B_y(x,y)=\frac{x}{\sqrt{x^2+y^2}}
\end{equation}
and we initialize $\mathcal{E}_{\rm cr}$ by setting high energy density in an annular sector:

\begin{equation}
    \mathcal{E}_{\rm cr}(t=0,x,y)=    \begin{cases}
      12, & {\rm for\ }0.5<r<0.7\ {\rm and\ } \phi<\pi/12 \\
      10, & {\rm other\ region}
    \end{cases}
\end{equation}
where $r\equiv\sqrt{x^2+y^2}$ and $\phi={\rm atan2}(y,x)$ is the angle between the positive $x$-axis and the ray from the origin to the point $(x,y)$. In this test we arrange both $\sigma_+$ and $\sigma_-$ constant regardless of the sign of $\nabla P_{\rm cr}$, so that the steady state flux equation becomes $\nabla P_{\rm cr}=-\left(\sigma_{+}+\sigma_{-}\right)F_{\rm cr}$ and $\mathcal{E}_{\rm cr}$ is subject to a traditional diffusion equation. The initial ${F}_{\rm cr}$ is set to zero. Under this circumstance, \cite{Pakmor16} has derived the analytical solution to compare with

\begin{equation}
\begin{aligned}
    \mathcal{E}_{\rm cr}(t,x,y)&=10+{\rm erfc}\left[\left(\phi-\frac{\pi}{12}\right)\frac{r}{D}\right]\\
    &-{\rm erfc}\left[\left(\phi+\frac{\pi}{12}\right)\frac{r}{D}\right]
    \end{aligned}
\end{equation}
where $D\equiv \sqrt{4t/3(\sigma_+ +\sigma_-)}$. We set $\sigma_+^L=\sigma_+^R=\sigma_-^L=\sigma_-^R=0.25$ so that $\sigma_+ + \sigma_-=1$. The initial gas density is  uniform with $\rho=1$, thus the Alfv\'en speed $v_A=1$. In this test, we turn off the MHD evolution; otherwise the magnetic tension of a circular field would squeeze gas inward, which then drags CRs and complicates the situation.

\begin{figure}
    \includegraphics[scale=0.276]{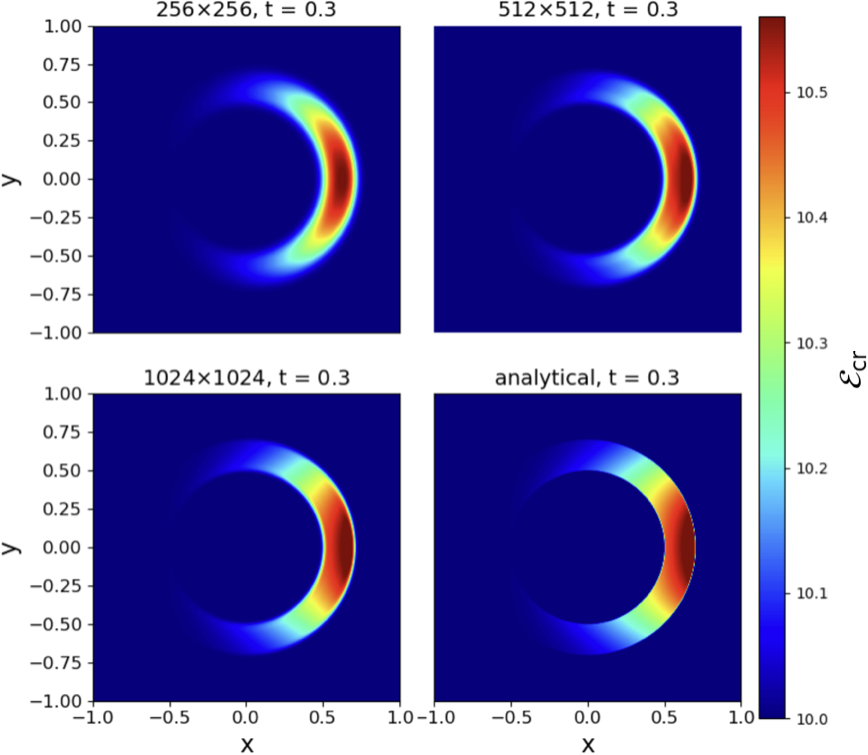}
    \caption{2D tests for the anisotropic transport of $\mathcal{E}_{\rm cr}$ along magnetic field lines on Cartesian grids, as described in \autoref{sec:aniso_flux}.  An initial patch with high CR energy density streams along circular background magnetic fields. The bottom right panel shows the analytical solution while other three panels demonstrate numerical results with different spatial resolutions at the same time.}
    \label{fig:circB}
\end{figure}

Our numerical results are compared with the analytical solution (bottom right) in \autoref{fig:circB}. All simulations exhibit the expected anisotropic transport. While lower-resolution runs show increased numerical diffusion, the $1024 \times 1024$ run agrees closely with the analytical solution. The spatial convergence here is consistent with the corresponding test in \cite{JiangOh2018}.

\subsection{CR Streaming in Spherical Coordinates}\label{subsec:sphCRsttest}
Our CR module also supports curvilinear coordinate systems such as spherical and cylindrical coordinates. In this section we show a test in a 3D spherical coordinate.

We use a CR streaming problem similar to the one described in \autoref{subsec:1Dstreaming}, except we now run it on a spherical grid with $512\times 32\times 8$ cells in the $r$, $\theta$ and $\phi$ directions, respectively. This time, we initialize the $\mathcal{E}_{\rm cr}$ profile by setting $\mathcal{E}_{\rm cr,0}=2$ in the central core $r<r_{\rm core}\ (r_{\rm core}=1.0)$, and 0 in the rest of the simulation domain. According to \autoref{eq:frontspeed}, the initial discontinuous front should propagate at the speed $v_{\rm front}=(1+4\widetilde{\sigma}_{0,\rm st})v_A/3\widetilde{\sigma}_{0,\rm st}$, where we set $\widetilde{\sigma}_{0,\rm st}=1$. For the purpose of this test we impose a monopole magnetic field so that the CR streaming is only along the radial direction, and we arrange the gas density accordingly to keep $v_A=1$ also a constant. Although a monopole magnetic field is unphysical, here we turn off the MHD evolution and the setup for gas only serves as a static platform.

We use outflow condition for both inner and outter radial boundaries, so that one can think of the central cavity of the domain as filled with the same value of $\mathcal{E}_{\rm cr}$ as the surroundings, and the central value can then be calculated by energy conservation as the system evolves:
\begin{equation}
    \mathcal{E}_{\rm cr}(t,r)=    \begin{cases}
      \mathcal{E}_{\rm cr,0}\left(\frac{r_{\rm core}}{r_{\rm core}+v_{\rm front}t}\right)^3, & r<r_{\rm core}+v_{\rm front}t\\
      0, & r>r_{\rm core}+v_{\rm front}t
    \end{cases}
\end{equation}

\begin{figure}
    \includegraphics[scale=0.28]{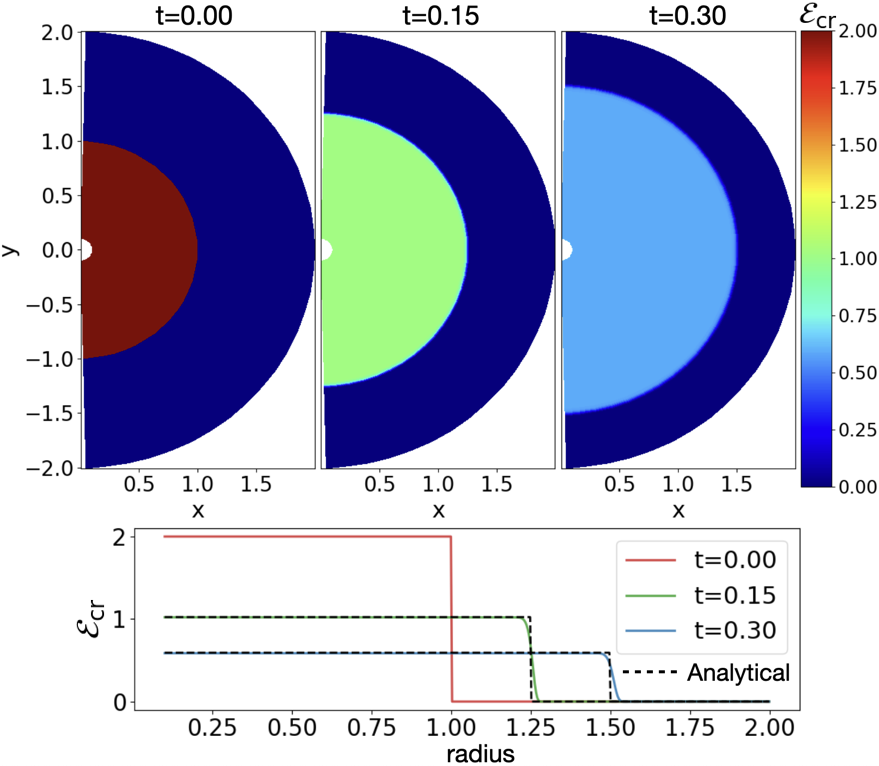}
    \caption{Numerical results of the CR streaming problem in a 3D spherical coordinate system, as described in \autoref{subsec:sphCRsttest}, with CR transport along the radial direction. {\bf{Top panels:}} $\mathcal{E}_{\rm cr}$ profiles shown in $r-\theta$ slices at different times. The boundary separating the central high-$\mathcal{E}_{\rm cr}$ region and the outer zero-$\mathcal{E}_{\rm cr}$ region should expand at a constant speed $v_{\rm front}$, with $\mathcal{E}_{\rm cr}$ being constant in the central region, whose value can be calculated through CR energy conservation. {\bf{Bottom panel:}} radial profiles of $\mathcal{E}_{\rm cr}$ compared with the theoretical expectation. We see they agree well and our code can properly handle the sharp discontinuity.}
    \label{fig:CRstream_sph}
\end{figure}

In \autoref{fig:CRstream_sph}, we compare this theoretical expectation with our numerical results. We see the simulation works well in the spherical coordinates, and \autoref{fig:CRstream_sph} also demonstrates that our code can properly deal with a sharp discontinuity as long as the stability criterion for the timestep \autoref{eq:stabcriterion} is satisfied.

\subsection{CRPAI in Cylindrical Coordinates}\label{subsec:CRPAItest}

In this subsection, we design a test problem to evaluate terms involving CR pressure anisostropy $\Delta P_{\rm cr}$, within a cylindrical coordinate. To ensure zero divergence, we prescribe magnetic fields $B(R)=1/R$ where $R$ is the cylindrical radius. The background MHD gas density is set uniformly to $\rho_{\rm gas}=1$, resulting in the Alfv\'en speed $v_{A}= 1/R$. Throughout the simulations, we freeze the MHD part so that the initial configuration is fixed.

We initialize $\mathcal{E}_{\rm cr}$ with a central peak, prompting CRs to stream outward along field lines. Consequently, as CRs traverse regions with varying magnetic field strengths, adiabatic invariance induces CR pressure anisotropy, activating the CRPAI. Following \autoref{eq:CRPAIscattpres} and \autoref{eq:DeltaPcrpres}, in this scenario we prescribe the scattering coefficients and $\Delta P_{\rm cr}$ as:
\begin{equation}
    \begin{aligned}
            &\Delta P_{\rm cr}=\frac{1}{\widetilde{\sigma}_{0,\rm aniso}}\frac{v_A}{c}P_{\rm cr}\\
        &\sigma_+^R=\sigma_-^L=\frac{\widetilde{\sigma}_{0,\rm aniso}}{2}\frac{1}{v_A c}\left|\frac{\dot{B}}{B}\right|\\
        &\sigma_+^L=\sigma_-^R=0
    \end{aligned}
\end{equation}
where the dimensionless constant $\widetilde{\sigma}_{0,\rm aniso}$ encapsulates the environmental parameters. In this test, we turn on the $\mathcal{E}_{\rm cr}$ source term; otherwise the scattering coefficients in front of $\Delta P_{\rm cr}$ in the $\vec{F}_{\rm cr}$ source would cancel, allowing $\Delta P_{\rm cr}$ to influence evolution only through the CR energy source term.

For this test, we arrive at the following CR equations:
\begin{equation}
\begin{aligned}
    &\frac{\partial \mathcal{E}_{\rm cr}}{\partial t}+\frac{1}{R}\frac{\partial (RF_{\rm cr})}{\partial R}\\
    &=\frac{4}{3}(\sigma_+^R+\sigma_-^L)\mathcal{E}_{\rm cr}v_A^2 - (\sigma_+^R+\sigma_-^L)v_A c\Delta P_{\rm cr}\\
    &=\left(\frac{4\widetilde{\sigma}_{0,\rm aniso}-1}{3}\right)\frac{1}{B}\left|\frac{dB}{dR}\right|\mathcal{E}_{\rm cr}v_A
    \end{aligned}
\end{equation}
and 
\begin{equation}
\begin{aligned}
    &\frac{1}{V_m^2}\frac{\partial F_{\rm cr}}{\partial t}+\frac{1}{3}\frac{\partial \mathcal{E}_{\rm cr}}{\partial R}=-(\sigma_+^R+\sigma_-^L)F_{\rm cr}\\
    &=-\frac{\widetilde{\sigma}_{0,\rm aniso}}{v_A}\frac{1}{B}\left|\frac{dB}{dR}\right|F_{\rm cr}.
    \end{aligned}
\end{equation}

Generally speaking, the reference solution of $\mathcal{E}_{\rm cr}$ is a series of Bessel functions, and we leave the derivation in \autoref{append:cylindricalCRPAI}. For simplification, we further set $\widetilde{\sigma}_{0,\rm aniso}=0.25$ so that the CR energy source term vanishes. Physically it means we artificially balance Fermi acceleration with energy loss through CR heating. In this case, one reference solution of $\mathcal{E}_{\rm cr}$ is $\mathcal{E}_{\rm cr}(t,R)=J_0(\sqrt{3}R/2)e^{-t}$, where $J_0$ is the zeroth order Bessel function of the first kind. We use this theoretical prediction as initial condition, and prescribe it in the ghost zones, expecting an overall exponential decay $e^{-t}$ in the simulations.

\begin{figure}
    \includegraphics[scale=0.265]{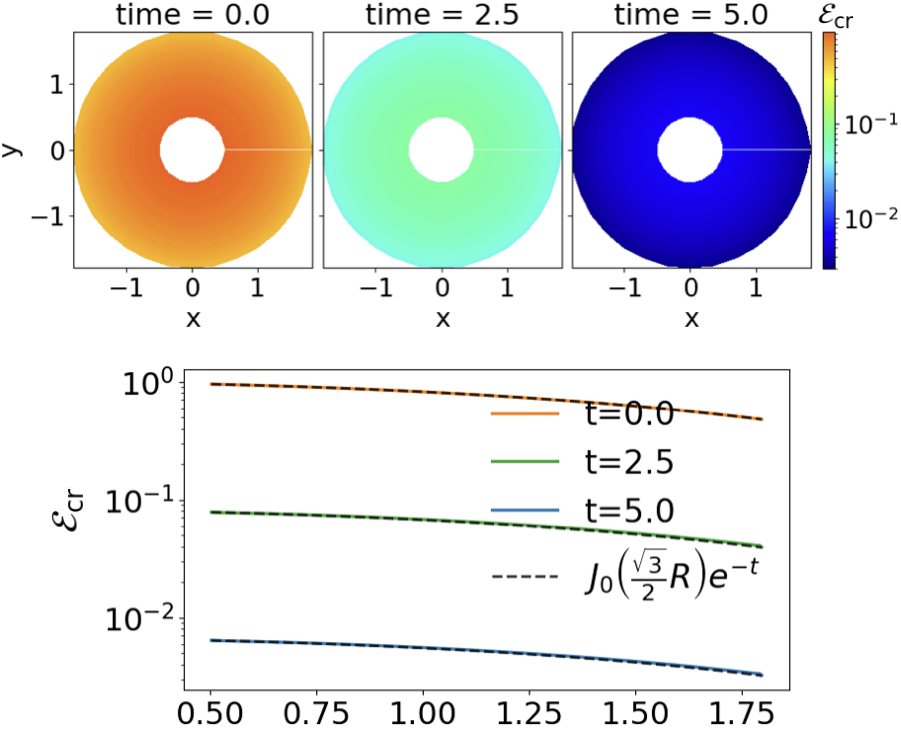}
    \caption{Evolution of $\mathcal{E}_{\rm cr}$ in a 2D cylindrical coordinate, where the CR scattering is subject to CRPAI, as described in \autoref{subsec:sphCRsttest}. {\bf Top panels:} 2D profiles of $\mathcal{E}_{\rm cr}$ at different times (title). We see the system is well symmetric in the angular direction. {\bf Bottom panel:} radial profiles of $\mathcal{E}_{\rm cr}$ at the same moments as in the top panels. Simulations well match the expected exponential decay.}
    \label{fig:CRPAI_cylin}
\end{figure}

In the top panels of \autoref{fig:CRPAI_cylin}, we present the 2D profiles of $\mathcal{E}_{\rm cr}$ at different times, illustrating that angular symmetry is preserved throughout the simulations. The bottom panel compares the numerical results with the theoretically predicted exponential decay and shows excellent agreement. Combined with the maintained cylindrical symmetry, these results validate our implementations of both the explicit $\Delta P_{\rm cr}$ source terms and our cylindrical coordinates.

Given the current limited understanding of CRPAI, we emphasize that this problem setup is intentionally idealized and serves solely to verify code accuracy. A more realistic physical treatment will require further investigation.

\section{Numerical Tests of the full CR-MHD system}\label{sec:5}

We now turn on the CR energy and momentum feedback to the MHD gas, validating our code for the coupled CR-MHD system. In doing so, we conduct the first linear analysis of the full two-moment CR-MHD wave system. The problem setup and full derivation/analysis are provided in \autoref{append:waves}. Early investigations of CR-MHD waves by \cite{Begelman1997} have revealed the presence of CR acoustic instability, which was revisited numerically by \cite{Tsung2022} recently. Our analysis focuses on the waves in a uniform fluid and CR background, but assumes the presence of a weak background $\nabla P_{\rm cr}$ to set the scattering rate, which is taken to be constant. Major highlights include:
\begin{itemize}
    \item In general, a CR–MHD system has four wave branches. In the weak-scattering limit, where CRs and MHD gas are decoupled, they reduce to forward- and backward-propagating acoustic waves in the thermal and CR fluids, respectively. As the scattering rate increases toward the fully coupled regime, the system supports forward and backward sound waves of the combined fluid, along with a single CR-modified acoustic mode. A fourth mode propagates backward at the Alfv\'en speed and is rapidly damped.

    \item The CR-modified acoustic wave can become unstable when $v_A > C_s$, consistent with previous findings by \cite{Begelman1997} and \cite{Tsung2023}.
\end{itemize}

\begin{table*}[h!]
\begin{tabular}{c c c c c c c c c c} 
 \hline
Identifier & $\delta \rho/\rho_0$ & $v_{A,0}$ & $C_s$ & $C_{\rm cr}$ & $V_m$ & $k$ & ${\rm Re}(\widetilde{\omega})$ & ${\rm Im}(\widetilde{\omega})$ & $\widetilde{\sigma}$\\ [0.5ex] 
 \hline\hline
 CR-MHD acoustic wave & $10^{-3}$ & 0.01 & 0.1 & 1 & 100 & 0.06 & 100.2 & $-10.49$ & 0.02\\ 
damping CR-modified acoustic wave & $10^{-3}$ & 0.01 & 0.1 & 1 & 100 & 5.25 & 9.60 & $-2.72$ & $2\times 10^{-4}$\\ 
growing CR-modified acoustic wave & $10^{-3}$ & 0.1 & 0.01 & 1 & 1000 & 5.82 & 0.32 & $0.14$ & 0.002\\ 
rapidly damping wave & $10^{-8}$ & 0.1 & 0.01 & 1 & 1000 & 10.00 & $-1.01$ & $-10^5$ & 0.001\\ 
 \hline
\end{tabular}
\caption{Parameters in the simulations of linear waves in the CR-MHD system. Column 1: panel identifier. Column 2: relative magnitude of the initial gas density perturbation (background gas density $\rho_0=10$). Column 3-6: characteristic speeds, which constrain the background magnetic fields, thermal pressure, and CR energy densities. Column 7: wave vector $k$, which is equal to $2\pi/L_{\rm box}$ since we keep the simulation box size equal to one wavelength. Column 8-9: real and imaginary parts of the normalized wave frequency $\widetilde{\omega}\equiv\omega/kv_{A,0}$. Column 10: Normalized scattering coefficient $\widetilde{\sigma}\equiv \sigma v_{A,0}/k$, where we fix the physical scattering coefficient $\sigma=0.1$. These data also correspond to the $\star$ marks in \autoref{fig:A1}.}
\label{table:1}
\end{table*}


We choose a few representative cases of the CR-MHD waves for our tests, with simulation parameters listed in \autoref{table:1}. We set the uniform background according to the parameters $v_{A,0}\equiv B/\sqrt{\rho_0}$ (Alfv\'en speed), $C_s\equiv\sqrt{\gamma_{\rm gas}P_{\rm therm,0}/\rho_0}$ (thermal sound speed) and $C_{\rm cr}\equiv\sqrt{\gamma_{\rm cr}P_{\rm cr,0}/\rho_0}$ (CR sound speed), where we fix the background gas density $\rho_0=10$ and gas velocity $u_0=0$. The relative amplitude of the gas density perturbation is specified by $\delta\rho /\rho_0$ in the table, and the remaining quantities are determined accordingly by \autoref{eq:eigenu}-\autoref{eq:eigenFcr}.
In each simulation, we initialize the perturbations with planar waves, expecting them to evolve as predicted by the real parts of ${\rm exp}[i(kx-\omega t)]$. Here $k\equiv2\pi/L_{\rm box}$ is the wave vector and we adjust it by changing the simulation box size $L_{\rm box}$, which is set equal to the wavelength. We normalize the wave frequency as $\widetilde{\omega}\equiv \omega/k v_{A,0}$ in the table, so that ${\rm Re}(\widetilde{\omega})$ represents the wave speed in the units of $v_{A,0}$ while ${\rm Im}(\widetilde{\omega})$ represents the normalized damping or growth rates. 

\begin{figure*}
\centering
\includegraphics[scale=0.47]{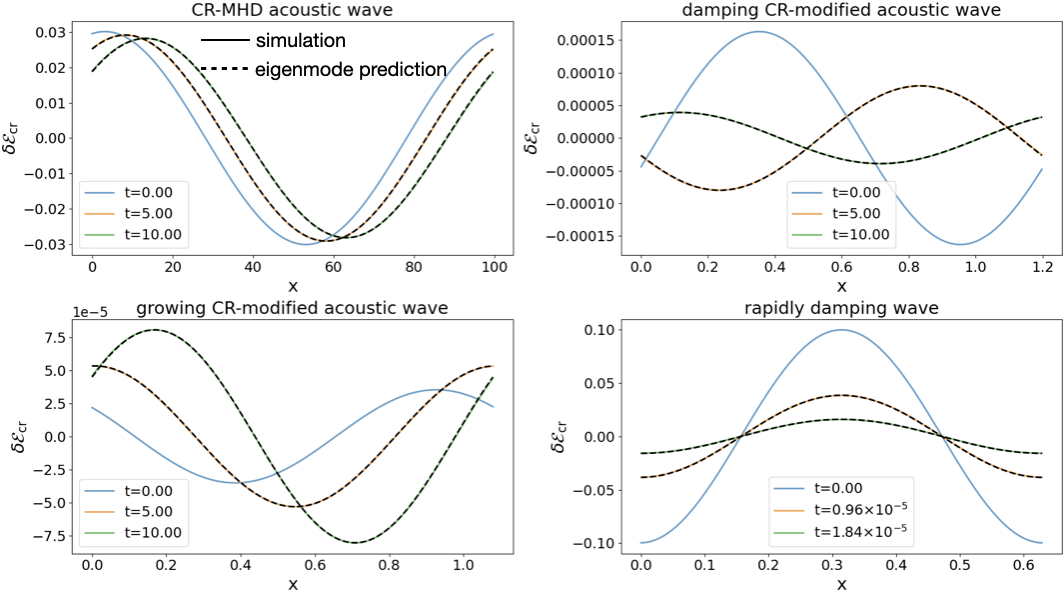}
    \caption{Validation of code performance in full CR-MHD systems. Numerical results (solid lines) and predictions from linear perturbations (dashed lines) in the coupled CR-MHD system are shown. Each panel tracks the evolution of an initially sinusoidal perturbation $\delta \mathcal{E}_{\rm cr}$ in four different situations: {\bf Top left:} Acoustic wave in the combined fluid where strong scattering tightly couples CRs and MHD. The sound speed is $C_{\rm cr}\equiv \sqrt{\gamma_{\rm cr}P_{\rm cr,0}/\rho_0}$ because $P_{\rm cr}$ dominates over other pressure components and the inertia is mostly provided by gas density $\rho_0$; {\bf Top right:} classic MHD acoustic wave (wave speed $=C_s$), weakly damped by CR interactions. {\bf Bottom left:} MHD acoustic wave,  driven to growing amplitude by CR-acoustic instability; {\bf Bottom right:} Rapidly damped Alfv\'en wave (wave speed $=v_{A,0}$
 ), decaying to machine-precision level so quickly that we only show results after a few steps. Specific parameters for each wave are listed in \autoref{table:1} and marked $\star$ in \autoref{fig:A1}, which maps wave branches across scattering regimes. Simulations demonstrate excellent agreement with theory, confirming code accuracy in diverse dynamical environments.}
    \label{fig:fulltest}
\end{figure*}

\autoref{fig:fulltest} compares the numerical evolution of $\delta\mathcal{E}_{\rm cr}$ ($\mathcal{E}_{\rm cr}$ perturbations) with linear wave predictions across four distinct dynamical regimes: In the top left panel, the scattering tightly couples CRs and the MHD gas as a combined fluid, with $\mathcal{E}_{\rm cr}$ dominating over other energy components and $\rho_0$ providing inertia, such that the wave speed is determined by $C_{\rm cr}$. The top right and bottom left panels illustrate CR-modified acoustic waves, where the wave speed equals $C_s$; the former exhibits damping, while the latter shows growth, depending on the ratio $C_s/v_{A,0}$. Finally, the bottom right panel presents a case with an extremely high damping rate, where the perturbation decays rapidly to the machine-precision level, so we only show the early-stage evolution. We see our module
is fully capable of capturing the expected wave evolutions across all the four dynamical regimes.

\section{Discussion}\label{sec:6}
In this section we compare our framework with previous ones, comments on future applications and possible extensions, and remark on caveats.

\subsection{Links to Previous Two-Moment Frameworks}

\subsubsection{Comparison with \cite{JiangOh2018}}

The numerical scheme in \cite{JiangOh2018} is 
conceived within the context of waves generated by CRSI, such that 
only Alfv\'en waves along one direction can be present (either along or opposite to the local $B$ field, but not both), with the CR pressure anisotropy $\Delta P_{\rm cr}$ not included. We show that in this special case, our equations can reduce to the two-moment scheme in \cite{JiangOh2018}.

If Alfv\'en waves only propagate along one direction, then only one of $\sigma_+$ and $\sigma_-$ is non-zero, depending on the sign of $\vec{b}\cdot\nabla P_{\rm cr}$. We denote the non-zero scattering coefficient just as $\sigma$, and define streaming velocity $\vec{v}_{\rm st}\equiv-{\rm sgn}(\vec{b}\cdot\nabla P_{\rm cr})\vec{v}_A$, then our CR equations can be written as:

\begin{equation}\label{eq:JOhCRenergy}
    \frac{\partial \mathcal{E}_{\rm cr}}{\partial t}+\nabla\cdot\vec{F}_{\rm cr}=-\sigma[\vec{F}_{\rm cr}-(\mathcal{E}_{\rm cr}+P_{\rm cr})(\vec{u}+\vec{v}_{\rm st})]\cdot(\vec{u}_{\parallel}+\vec{v}_{\rm st})
\end{equation}

\begin{equation}\label{eq:JOhCRflux}
    \frac{1}{V_m^2}\frac{\partial \vec{F}_{\rm cr}}{\partial t}+\nabla\cdot \mathbf{P}_{\rm cr}=-\sigma[\vec{F}_{\rm cr}-(\mathcal{E}_{\rm cr}+P_{\rm cr})(\vec{u}+\vec{v}_{\rm st})]\cdot\vec{b}\vec{b}.
\end{equation}
We note that \cite{JiangOh2018} use large perpendicular scattering coefficients to constrain $\vec{F}_{\rm cr}-(\mathcal{E}_{\rm cr}+P_{\rm cr})\vec{u}$ aligned with $\vec{b}$, so that the Lorentz force terms vanish.

Next, assuming steady-state CR flux (set $\partial \vec{F}_{\rm cr}/\partial t=0$), we dot equation \autoref{eq:JOhCRflux} with $(\vec{u}+\vec{v}_{\rm st})$ and plug it back into equation \autoref{eq:JOhCRenergy}, obtaining:

\begin{equation}\label{eq:YFJCRenergy}
    \frac{\partial \mathcal{E}_{\rm cr}}{\partial t}+\nabla\cdot\vec{F}_{\rm cr}=(\vec{u}+\vec{v}_{\rm st})\cdot(\nabla\cdot\mathbf{P}_{\rm cr})
\end{equation}
On the other hand, moving the $\vec{v}_{\rm st}$ term in \autoref{eq:JOhCRflux} to the LHS provides (note that ignoring $\Delta P_{\rm cr}$ gives $\nabla\cdot\mathbf{P}_{\rm cr}=\nabla P_{\rm cr}$):

\begin{equation}
\begin{aligned}
    -\vec{b}&\cdot\left[\frac{1}{V_m^2\sigma}\frac{\partial \vec{F}_{\rm cr}}{\partial t}+\frac{\nabla P_{\rm cr}}{\sigma}+(\mathcal{E}_{\rm cr}+P_{\rm cr}){\rm sgn}(\vec{b}\cdot\nabla P_{\rm cr})v_A\right]\\
    &=-\vec{b}\cdot\left\{ \frac{1}{V_m^2\sigma}\frac{\partial\vec{F}_{\rm cr}}{\partial t} +\nabla P_{\rm cr}\left[\frac{1}{\sigma}+\frac{(\mathcal{E}_{\rm cr}+P_{\rm cr})\vec{v}_A}{\left|\vec{b}\cdot\nabla P_{\rm cr}\right|}\right]\right\}\\
    &=\vec{b}\cdot\left[\vec{F}_{\rm cr}-(\mathcal{E}_{\rm cr}+P_{\rm cr})\vec{u}\right]
    \end{aligned}
\end{equation}

\cite{JiangOh2018} then defined an effective scattering coefficient $\sigma^*$ as

\begin{equation}\label{eq:YFJscattering}
{\sigma^*}^{-1}=\sigma^{-1}+\frac{(\mathcal{E}_{\rm cr}+P_{\rm cr})v_A}{\left|\vec{b}\cdot\nabla P_{\rm cr}\right|}
\end{equation}
so that the CR flux equation can be modified as

\begin{equation}\label{eq:YFJCRflux}
    \frac{1}{{V}_m'^2}\frac{\partial\vec{F}_{\rm cr}}{\partial t}+\nabla \cdot\mathbf{P}_{\rm cr}=-\sigma^*\left[\vec{F}_{\rm cr}-(\mathcal{E}_{\rm cr}+P_{\rm cr})\vec{u}\right]
\end{equation}
where $V_m'^2=\sigma V_m^2/\sigma^*>V_m^2$ is just an increasing modification of reduced speed of light, and thus should not alter the outcome.

\autoref{eq:YFJCRenergy} and \autoref{eq:YFJCRflux} together with the definition \autoref{eq:YFJscattering} are identical to the equation (5) and (10) in \cite{JiangOh2018}, provided $\vec{F}_{\rm cr}$ is dominantly aligned with $\vec{b}$ with $\Delta P_{\rm cr}=0$. We thus can include their equations as a special case of our formulation, where the CR transport is dominated by CRSI and the effect of CR pressure anisotropy is not included. 

\subsubsection{Comparison with \cite{Thomas19}}

The formulation in \cite{Thomas19} adopts a first-principles starting point similar to ours. One key formal difference is that, by truncating at the first-order Legendre polynomial to model drift anisotropy in the CR distribution, their approach precludes CR pressure anisotropy modes and the associated microphysics. It is therefore straightforward to reduce to their formulation simply by setting $\Delta P_{\rm cr}=0$ in our equations, which yields:

\begin{equation}\label{eq:Ecr_Pfrommer}
\begin{aligned}
    &\frac{\partial \mathcal{E}_{\rm cr}}{\partial t}+\nabla\cdot\vec{F}_{\rm cr}=\\
    &-\sigma_+\left[\vec{F}_{\rm cr}-(\mathcal{E}_{\rm cr}+P_{\rm cr})(\vec{u}+\vec{v}_A)\right]\cdot(\vec{u}_{\parallel}+\vec{v}_A)\\
    &-\sigma_-\left[\vec{F}_{\rm cr}-(\mathcal{E}_{\rm cr}+P_{\rm cr})(\vec{u}-\vec{v}_A)\right]\cdot(\vec{u}_{\parallel}-\vec{v}_A)\\
    &+\frac{A\Omega_{\rm cr}}{c^2}\left(\vec{F}_{\rm cr}\times\vec{b}\right)\cdot\vec{u}
    \end{aligned}
\end{equation}

\begin{equation}\label{eq:Fcr_Pfrommer}
\begin{aligned}
    &\frac{1}{V_m^2}\frac{\partial \vec{F}_{\rm cr}}{\partial t}+\nabla\cdot\mathbf{P}_{\rm cr}=\\
    &-\sigma_+\left[\vec{F}_{\rm cr}-(\mathcal{E}_{\rm cr}+P_{\rm cr})(\vec{u}+\vec{v}_A)\right]\cdot\vec{b}\vec{b}\\
    &-\sigma_-\left[\vec{F}_{\rm cr}-(\mathcal{E}_{\rm cr}+P_{\rm cr})(\vec{u}-\vec{v}_A)\right]\cdot\vec{b}\vec{b}\\
    &+\frac{A\Omega_{\rm cr}}{c^2}\left[\vec{F}_{\rm cr}-(\mathcal{E}_{\rm cr}+P_{\rm cr})\vec{u}\right]\times\vec{b}
    \end{aligned}
\end{equation}

These two equations are identical to their lab-frame CR equations (see equation E13 and E14 in \cite{Thomas19}, or equation 31 and 32 in \cite{Ruszkowski23}).

Another important conceptual distinction lies in how scattering coefficients $\sigma_{\pm}^{L,R}$ are determined. \citet{Thomas19} co-evolve the Alfvén wave subsystem dynamically alongside CR-MHD equations to obtain instantaneous $\sigma_{\pm}^{L,R}$ associated with Alfv\'en wave intensities, rather than prescribing a local steady-state balance as in our work. Their approach thus naturally accommodates regimes such as “Alfvén-wave dark regions” \citep{Thomas2023}, where $\nabla P_{\rm cr}$ is nearly perpendicular to the magnetic field, and the assumption of rapid wave growth/damping balance no longer holds due to suppressed gyro-resonant instabilities.

In typical ISM conditions, however, wave growth from gyro-resonant instabilities (e.g., CRSI and CRPAI) saturates on timescales much shorter than the dynamical timescale resolved in large-scale simulations, with the wave subsystem converging toward QLT predictions. More subtly, our steady-state scattering coefficients can be directly calibrated against MHD-PIC simulations \citep[e.g.,][]{Bai2022,Sun2024}, capturing deviations from QLT predictions. By comparison, calibrating the dynamically evolving wave subsystem is less straightforward. We therefore view the two methods as complementary, with the steady-state framework providing a practical and robust tool for astrophysical applications where rapid kinetic equilibration is expected.



\subsection{Limitations and Future Work}

Our CR hydrodynamics provides a versatile foundation for modeling CR transport mediated by resonant Alfv\'en wave scattering, in principle accommodating arbitrary wave configurations. However, in practice, wave amplitudes are rarely known a priori, necessitating self-consistent modeling that balances wave driving (e.g., CRSI, CRPAI, or turbulent cascades) against damping mechanisms (e.g., ion-neutral friction, Landau damping). While the CRSI-dominated regimes has been extensively studied, CRPAI remains underexplored, and no unified theory currently addresses combined driving/damping processes in multiphase or turbulent systems. This ambiguity in scattering coefficients poses a significant barrier to advancing predictive CR-MHD models. Further complexities arise from non-wave scattering transport, such as the strong scattering due to magnetic field-line reversals at intermittencies in MHD turbulence \citep{Kempski2023,Kempski2025}. These transport processes, in regimes where wave-mediated transport is subdominant, are currently beyond the scope of our framework. Addressing these theoretical gaps and extending the framework to incorporate non-wave scattering mechanisms are critical next steps reserved for future studies.

While this work focuses primarily on $\sim$GeV CRs which dominate CR energetics, a complete CR feedback model should account for their energy-dependent astrophysical roles. For instance, low-energy CRs govern ionization and molecular chemistry, whereas high-energy CRs drive dynamical processes like galactic wind acceleration. These distinct effects necessitate energy-resolved modeling, which requires implementation of multiple CR species \citep[e.g.][]{Hopkins+22a,Girichidis+20,  Armillotta_2025,Linzer_2025} because the dominant CR transport mechanisms vary with CR energy. Extending the framework to a multi-energy-bin version also enables direct comparison with observed CR spectra. Although our current implementation includes only a single CR species, its modular architecture enables seamless integration of additional energy-resolved components. This extensibility lays the groundwork for a unified multi-energy CR module, capable of capturing the full spectrum of CR astrophysical impacts while facilitating comparison with observations.

Finally, we note that our formulation also incorporates the CR pressure anisotropy $\Delta P_{\rm cr}$ in the advection term $\nabla\cdot\mathbf{P}_{\rm cr}$, and we have provided the corresponding numerical interface. However, we have not conducted dedicated tests, as the underlying physics remains insufficiently investigated.

In this work, we focus on idealized tests without introducing physical units. These tests are designed to isolate physical effects and verify the numerical implementation and performance of our CR module, as demonstrated in this work. Future efforts will apply this framework to realistic astrophysical environments, with such applications presented in forthcoming work. 

\section*{Acknowledgements}

We are especially grateful to Pinghui Huang for sharing his multifluid dust module in \textit{Athena}++ as a template for our implementation. We also thank Minghao Guo, Zitao Hu, and Shengtang Wang for their helpful suggestions during the development process, and Phil Hopkins, Xiaochen Sun, Shuzhe Zeng, Suoqing Ji, Philipp Kempski, Chang-Goo Kim, Hanjue Zhu, and Alexander Wagner for valuable scientific discussions and suggestions. We thank the anonymous referee for thoughtful and constructive comments that helped improve the manuscript.

This work is supported by the National Science Foundation of China under grant No. 12325304, 12342501, and in part by grant 510940 from the Simons Foundation to ECO.

\appendix

\section{CR-MHD waves}\label{append:waves}

In this appendix we calculate characteristic waves in a CR-MHD system. This topic has been studied by \cite{Begelman1997,Tsung2022}, which focused on unstable waves arising from CR-modified acoustic instabilities. Here we perform a separate derivation mainly serving two objectives: (1) setting the stage for our full CR-MHD system tests in \autoref{sec:5}, where we validate our numerical solutions against analytical eigenmode predictions; (2) determining the maximum wave speeds for the implementation in our HLLE Riemann solver.

For simplicity we only consider a 1D situation where all perturbations are longitudinal. In this case $\nabla\cdot \vec{B}=0$ implies $B={\rm const}$, and the induction equation then becomes redundant. We further replace the gas energy equation by an barotropic equation of state (EOS), which simplifies the system by discarding the entropy wave while retaining the physics of an adiabatic gas for linear problems. We assume the presence of a background $\nabla P_{\rm cr}$ directed downstream along the magnetic field, such that only $\sigma_+$ is non-zero while $\sigma_-=0$. This background gradient $\nabla P_{\rm cr}$ is taken to be sufficiently weak that the background CR pressure, $P_{\rm cr,0}$, can be treated as spatially constant in our analysis—yet still large enough that it is not reversed by linear perturbations. This approximation is valid when the wavelengths of interest are much smaller than the characteristic length scale of the $P_{\rm cr}$ gradient. As is standard practice, all other physical quantities are assumed to have uniform background states. The equations we linearize are thus as follows:
\begin{equation}
    {\rm Mass\ conservation}: \frac{\partial \rho}{\partial t}+\frac{\partial(\rho{u})}{\partial x}=0
\end{equation}

\begin{equation}
    {\rm Gas\ momentum}:  \rho\frac{\partial u}{\partial t}+\rho u\frac{\partial u}{\partial x}+\frac{\partial P_{\rm therm}}{\partial x}=\sigma\left[F_{\rm cr}-4(u+v_A)P_{\rm cr}\right]
\end{equation}

\begin{equation}
    {\rm Barotropic\ EOS}: P_{\rm therm}={\rm const}\times\rho^{\gamma}
\end{equation}

\begin{equation}
    {\rm CR\ energy}: \frac{\partial \mathcal{E}_{\rm cr}}{\partial t}+\frac{\partial F_{\rm cr}}{\partial x}=-\sigma\left[F_{\rm cr}-4(u+v_A)P_{\rm cr}\right](u+v_A)
\end{equation}

\begin{equation}
    {\rm CR\ flux}: 
    \frac{1}{V_m^2}\frac{\partial F_{\rm cr}}{\partial t}+\frac{\partial P_{\rm cr}}{\partial x}=-\sigma\left[F_{\rm cr}-4(u+v_A)P_{\rm cr}\right]
\end{equation}
where $\sigma$ is a constant, and all other variables follow the same notation as in the main text.

Next, we decompose all quantities into background values plus plane wave perturbations, i.e., any quantity $X$ can be expressed as $X=X_0+\delta X e^{i(kx-\omega t)}$, where $X_0$ is the steady-state background and $|\delta X|\ll |X_0|$ is the small amplitude of perturbation (note that $\delta X$ is a complex amplitude that contains both magnitude and phase information). 
After subtracting the balanced background and neglecting higher-order terms, we obtain the following linearized equations:
\begin{equation}
    i\left(\omega-ku_0\right)\delta \rho-ik\rho_0 \delta u = 0  
\end{equation}

\begin{equation}
i\left(ku_0-\omega\right)\rho_0\delta u+ik\delta P_{\rm therm}=\sigma \rho_0 C_{\rm cr}^2\left[\frac{3}{4}\frac{\delta F_{\rm cr}}{P_{\rm cr,0}}-3\delta u+\frac{3}{2}\frac{v_{A,0}\delta \rho}{\rho_0}-3\left(u_0+v_{A,0}\right)\frac{\delta P_{\rm cr}}{P_{\rm cr,0}}\right]
\end{equation}

\begin{equation}
    \delta P_{\rm therm}=C_s^2\delta \rho
\end{equation}

\begin{equation}
\begin{aligned}
    -3i\omega \delta P_{\rm cr} + ik\delta F_{\rm cr}=&-\sigma \rho_0 C_{\rm cr}^2\left[\frac{3}{4}\frac{\delta F_{\rm cr}}{P_{\rm cr,0}}-3\delta u+\frac{3}{2}\frac{v_{A,0}\delta \rho}{\rho_0}-3\left(u_0+v_{A,0}\right)\frac{\delta P_{\rm cr}}{P_{\rm cr,0}}\right]\left(u_0+v_{A,0}\right)&
    \end{aligned}
\end{equation}

\begin{equation}
    -\frac{i\omega}{V_m^2}\delta F_{\rm cr}+ik\delta P_{\rm cr}=-\sigma \rho_0 C_{\rm cr}^2\left[\frac{3}{4}\frac{\delta F_{\rm cr}}{P_{\rm cr,0}}-3\delta u+\frac{3}{2}\frac{v_{A,0}\delta \rho}{\rho_0}-3\left(u_0+v_{A,0}\right)\frac{\delta P_{\rm cr}}{P_{\rm cr,0}}\right]
\end{equation}
where $C_s^2\equiv \gamma_{\rm gas} P_{\rm therm,0}/\rho_0$ and $C_{\rm cr}^2\equiv \gamma_{\rm cr}P_{\rm cr,0}/\rho_0=4P_{\rm cr,0}/3\rho_0$ are separately gas sound speed and CR sound speed. As elsewhere, note that we are using magnetic units in which $v_A\equiv B/\sqrt{\rho}$.

For clarity we normalize everything by $v_{A,0}$ and $k$ to define a set of dimensionless parameters: $\widetilde{\omega}\equiv(\omega-ku_0)/kv_{A,0}$, 
$\widetilde{\sigma}\equiv \sigma v_{A,0}/k$,  $\widetilde{u}_0\equiv u_0/v_{A,0}$, $\widetilde{C}_s\equiv C_s/v_{A,0}$, $\widetilde{C}_{\rm cr}\equiv C_{\rm cr}/v_{A,0}$, and  $\widetilde{V}_m\equiv V_m/v_{A,0}$. 
The eigenstate relations are:
\begin{equation}\label{eq:eigenu}
    \frac{\delta u}{v_{A,0}} =\widetilde{\omega}  \frac{\delta\rho}{\rho_0},
\end{equation}
\begin{equation}\label{eq:eigenPcr}
\frac{\delta P_{\rm cr}}{\rho_0 v_{A,0}^2}=\frac{1}{3}\frac{\delta \mathcal{E}_{\rm cr}}{\rho_0 v_{A,0}^2}=\frac{\left[\widetilde{V}_m^2+(1+\widetilde{u}_0)(\widetilde{\omega}+\widetilde{u}_0)\right](\widetilde{\omega}^2-\widetilde{C}_s^2)}{\left[\widetilde{V}_m^2-3{(\widetilde{\omega}+\widetilde{u}_0)^2}\right]} \frac{\delta \rho}{\rho_0}, 
\end{equation}

\begin{equation}\label{eq:eigenFcr}
\frac{\delta F_{\rm cr}}{\rho_0 v_{A,0}^3}=\frac{\widetilde{V}_m^2\left(1+3\widetilde{\omega}+4\widetilde{u}\right)(\widetilde{\omega}^2-\widetilde{C}_s^2)}{\left[\widetilde{V}_m^2-3{(\widetilde{\omega}+\widetilde{u}_0)^2}\right]}\frac{\delta \rho}{\rho_0}.  
\end{equation}

After neglecting the terms scaling as $\mathcal{O}(C/V_m)$ ($C$ refers to any characteristic speed other than $V_m$, where $V_m$ is intentionally chosen to be much larger than all other speeds), we arrive at the following dispersion relation:

\begin{equation}
\label{eq:dispersion}
\frac{\widetilde{\omega}^4}{\widetilde{V}_m^2}+i\widetilde{\sigma}\widetilde{\omega}^3-\left(\frac{1}{3}+i\widetilde{\sigma}\right)\widetilde{\omega}^2-i\widetilde{\sigma}\left(\widetilde{C}_s^2+\widetilde{C}_{\rm cr}^2\right)\widetilde{\omega}+\frac{\widetilde{C}_s^2}{3}+i\widetilde{\sigma}\left(\widetilde{C}_s^2+\frac{\widetilde{C}_{\rm cr}^2}{2}\right)=0,
\end{equation}

The fact that $u_0$ is only implicitly contained in $\widetilde{\omega}$ reflects the system's Galilean invariance \footnote{The rigorous dispersion relation that preserves all terms is not fully Galilean invariant, because the CR particle speed is always treated as $c$ no matter in which frame. This relativistic feature essentially equates stringent Galilean invariance with an infinite speed of light, which is approximately true in the context of hydrodynamic-scale speeds.}.

\subsection{Portrait of waves in a CR-MHD system}\label{sec:CR-MHD_waves}

\begin{figure*}
\centering
\includegraphics[scale=0.43]{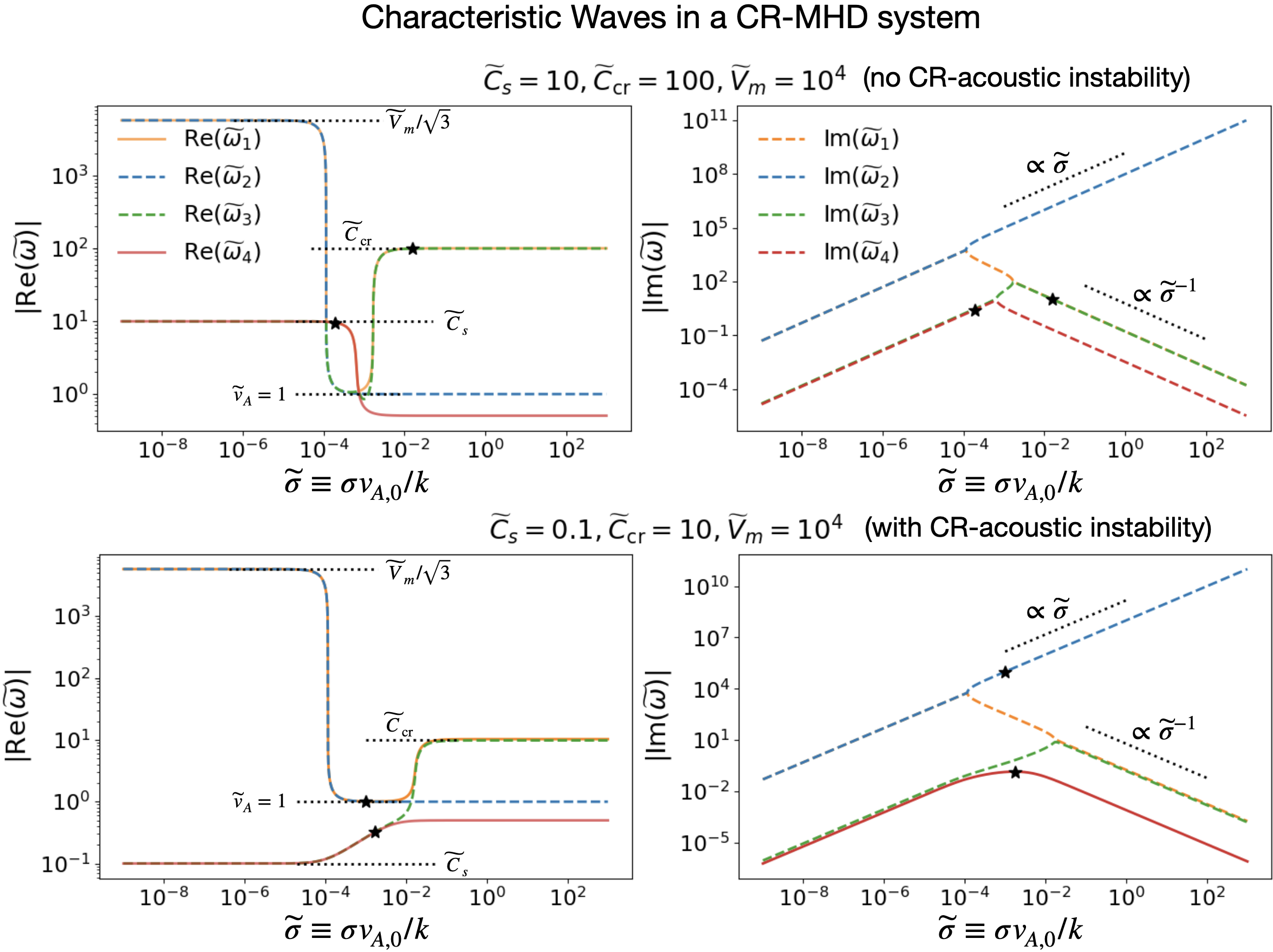}
    \caption{Roots of the dispersion relation \autoref{eq:dispersion} are plotted as functions of $\widetilde{\sigma}$, with fixed characteristic speeds (see panel titles). The left panels display the real parts, while right panels show the imaginary parts. Absolute values are displayed in logarithmic scale, with dashed/solid lines representing originally negative/positive values, respectively. {\bf Top panels} illustrate the stable regime $(\widetilde{C}_s \equiv C_s/v_{A,0}\gg 1)$, where all imaginary parts are negative, indicating no wave growth. {\bf Bottom panels} depict the unstable regime $(\widetilde{C}_s\ll 1)$, where the red branch has a positive imaginary part, signaling exponential growth. Since the real part of this branch approaches $\widetilde{C}_s$ in the weak scattering limit $(\widetilde{\sigma}\rightarrow0)$, it is identified as the CR-modified acoustic wave. In the real-part panels, we also mark (with horizontal dotted lines) other characteristic speeds to indicate asymptotic wave speeds. In the imaginary panels, the damping or growth rates are observed to asymptotically scale either linearly or inversely with $\widetilde{\sigma}$, as indicated with dotted lines. We use $\star$ to mark four points corresponding to the waves injected for testing our code in the full CR-MHD system in \autoref{sec:5}. Specific data for these $\star$ points are listed in \autoref{table:1}.}
    
    \label{fig:A1}
\end{figure*}

The quartic dispersion relation \autoref{eq:dispersion} generally has four roots corresponding to four wave branches. By fixing characteristic speeds $\widetilde{C}_s,\ \widetilde{C}_{\rm cr}$ and $\widetilde{V}_m$, we numerically solve for all roots and plot their dependence on $\widetilde{\sigma}$ in \autoref{fig:A1}. Points marked by $\star$ denote waves initialized in our numerical tests against eigenmode predictions (\autoref{sec:5}). For clarity in logarithmic scaling, we show absolute values of real and imaginary parts, with dashed lines indicating originally negative values. A similar dispersion analysis was also carried out by \cite{Thomas2021} for code verification.

We discuss two distinct regimes. The top panels demonstrate the real (left) and imaginary (right) parts of normalized wave frequency $\widetilde{\omega}$ in the case where $C_s\gg v_{A,0}$, and all imaginary parts are negative, indicating wave damping and no instability. However, when $C_s\ll v_{A,0}$, as in the bottom panels, one branch (red) exhibits a positive imaginary part (bottom right), signaling exponential growth of waves. The real part of this branch approaches ${\rm Re}(\omega)/k=C_s$ in the $\sigma v_{A,0}/k\rightarrow 0$ limit, and we thus identify it as the CR-acoustic instability described in \cite{Begelman1997} and \cite{Tsung2022}.

Both cases share asymptotic trends in real parts when $\sigma v_{A,0}/k\rightarrow 0$: $C_s$ and $V_m/\sqrt{3}$ emerge as two characteristic wave speeds. Physically, the reason is that when CRs fully decouple from the MHD gas, sound waves independently travel through the two fluids, with speeds being $C_s$ in MHD gas and $V_m/\sqrt{3}$ (which would physically be $c/\sqrt{3}$) in the CR fluid,
similar to the case of RHD. Increasing $\widetilde{\sigma}$ strengthens the CR-MHD coupling, driving wave speeds to deviate from their decoupled values. 
For very strongly coupled fluids at high $\widetilde{\sigma}$, a CR-MHD acoustic wave is evident, with wave speed ${\rm Re}(\omega)/k = {C}_\mathrm{cr}$ (i.e., the restoring force is provided by CR pressure when $C_\mathrm{cr}\gg C_s,v_{A,0}$).
For the imaginary parts, we see the damping/growth rates either increase or decrease linearly with $\widetilde{\sigma}$ in both cases (right panels). More detailed investigation of the physics behind these features is beyond the scope of this work and is left for future studies.

\subsection{Maximum wave speeds}

\begin{figure*}
\includegraphics[scale=0.386]{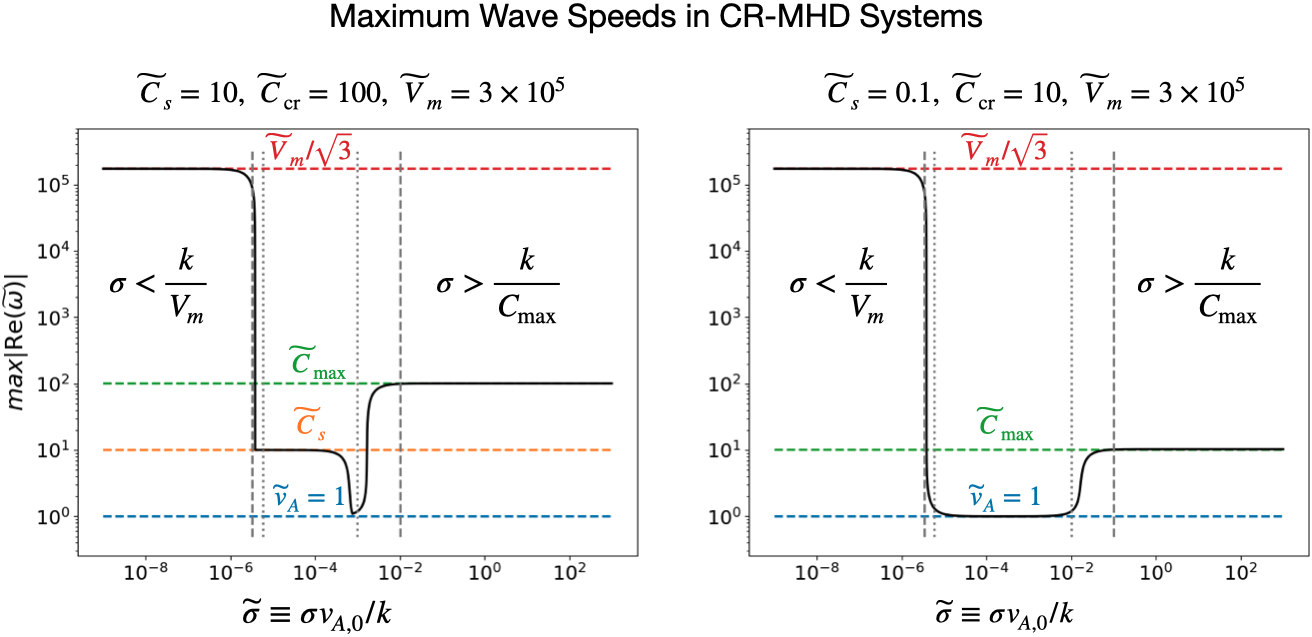}
    \caption{Maximum wave speeds in CR-MHD systems. Black solid lines indicate the maximum wave speeds in the left panels of \autoref{fig:A1}, and colored dashed lines represent other characteristic speeds for comparison. The left panel displays the stable regime, while the right panel shows the unstable regime. Gray dashed lines mark two key transitions: (1) $\widetilde{\sigma}=1/\widetilde{V}_m$, below which CRs and MHD are fully decoupled; (2) $\widetilde{\sigma}=1/\widetilde{C}_{\rm max}$, above which they are fully coupled. As $\widetilde{\sigma}$ increases, the maximum wave speed initially equals $V_m/\sqrt{3}$, representing relativistic gas sound speed in the decoupled regime, then it shifts to ${\rm max}(C_s, v_{A,0})$ in the intermediate region, and finally reaches $C_{\rm max}$ in the fully coupled regime. In the $C_s>v_{A,0}$ case (left panel), there is an additional interval where the wave speed is $v_{A,0}$ due to the CR-modified acoustic wave transitioning to a sub-Alfvénic state.
    For numerical stability, we aim to overestimate the maximum wave speed in our HLLE solver. Therefore, we employ two modified transition thresholds, indicated by gray dotted vertical lines: $\widetilde{\sigma}=\sqrt{3}/\widetilde{V}_m$ and $\widetilde{\sigma}=1/10\widetilde{C}_{\rm max}$, which effectively narrow the intermediate region.}
    \label{fig:A2}
\end{figure*}

The HLLE solver used in our CR module requires the maximum wave speeds of the CR–MHD system. Maximum wave speeds from the solutions shown in  \autoref{fig:A1} are plotted in \autoref{fig:A2}.
As pointed out in \autoref{sec:CR-MHD_waves}, in the weak scattering limit $\widetilde{\sigma}\equiv \sigma v_{A,0}/k\rightarrow 0$, the maximum wave speed is $V_m/\sqrt{3}$, i.e., the sound speed of the relativistic CRs (similar to RHD). This conclusion can be theoretically supported by solving the dispersion relation \autoref{eq:dispersion} while assuming $\widetilde{\sigma}\rightarrow 0$, which is:
\begin{equation}
    \widetilde{\omega}^4-\frac{\widetilde{V}^2_m}{3}\widetilde{\omega}^2+\frac{\widetilde{C}_s^2}{3}=0
\end{equation}
Considering $\widetilde{V}_m\gg\widetilde{C}_s$ in general, the two roots are separately $\widetilde{\omega}^2=\widetilde{V}_m^2/3$ and $\widetilde{C}_s^2$, corresponding to two separate sound speeds. 

In the strong scattering limit $\widetilde{\sigma}\rightarrow +\infty$, we keep only all the terms containing $\widetilde{\sigma}$ in equation \autoref{eq:dispersion}, yielding a cubic equation:

 \begin{equation}
     \widetilde{\omega}^3-\widetilde{\omega}^2-\left(\widetilde{C}_s^2+\widetilde{C}_{\rm cr}^2\right)\widetilde{\omega}+\left(\widetilde{C}_s^2+\frac{\widetilde{C}_{\rm cr}^2}{2}\right)=0
 \end{equation}
We estimate the maximum real part of the roots to be
\begin{equation}
\widetilde{C}_{\rm max}\equiv \sqrt{\widetilde{C}_s^2+\widetilde{C}_{\rm cr}^2+\widetilde{v}_{A,0}^2}=\frac{1}{v_{A,0}}\left(\frac{\gamma_{\rm gas}P_{\rm therm,0}+\gamma_{\rm cr}P_{\rm cr,0}+\gamma_mP_{B,0}}{\rho_0}\right)^{1/2}
\end{equation}
where $P_{B,0}\equiv B^2/2$ is the magnetic pressure and $\gamma_m=2$ is the corresponding adiabatic index. Physically, it can be understood by that, the strong scattering tightly couples MHD gas with CRs as a combined fluid, in which the sound speed is related to the total pressure acting on the gas inertia. In principle this approximation can be justified by solving the three roots of a cubic equation, but here we provide an order-of-magnitude observation. Define the cubic polynomial as $P(\widetilde{\omega})$, then we notice $P(+\infty)\rightarrow +\infty>0$ while $P(1)=-\widetilde{C}_{\rm cr}^2/2<0$, suggesting there must be one root $\widetilde{\omega}_*>1$. When $\widetilde{C}_{s}^2+\widetilde{C}_{\rm cr}^2\ll1$, the root thus satisfies $\widetilde{\omega}_*^3-\widetilde{\omega}_*^2\approx0$, giving $\widetilde{\omega}_*\approx 1$. On the other hand if $\widetilde{C}_{s}^2+\widetilde{C}_{\rm cr}^2\gg1$, by assuming $\widetilde{\omega}_*\gg1$ we find there is indeed a solution $\widetilde{\omega}_*^2=\widetilde{C}_{s}^2+\widetilde{C}_{\rm cr}^2\gg 1$, approximately satisfying the cubic equation. Therefore, by combining the three characteristic speeds as a sum of squares, $\widetilde{C}_{\rm max}$ automatically accommodates both situations.

\begin{figure*}
\includegraphics[scale=0.393]{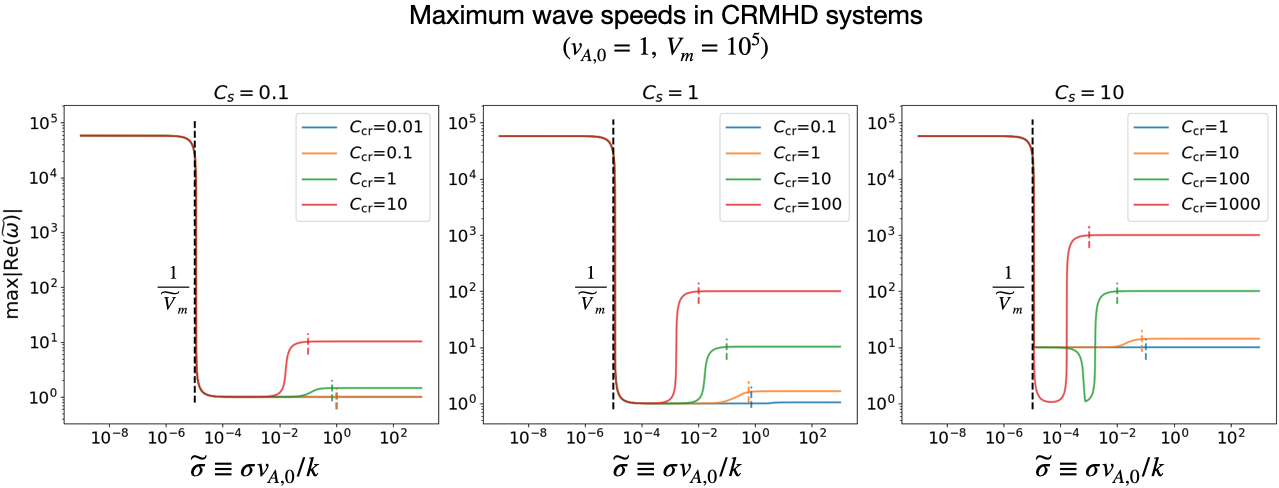}
    \caption{Maximum wave speeds in various CR-MHD systems with different parameters are shown here as a supplement to \autoref{fig:A2}. In these examples, the background Alfvén speed is fixed at $v_{A,0}=1$ (since it is the unit of other characteristic speeds in the dispersion relation), and the speed of light is set to $V_m=10^5$. From left to right, the thermal sound speed progressively increases from 0.1 to 10. Within each panel, different colors represent cases with various CR sound speeds $C_{\rm cr}$. Black dashed lines denote the first transition at $\widetilde{\sigma}=1/\widetilde{V}_m$, while colored dashed lines indicate corresponding second transitions at $\widetilde{\sigma}=1/\widetilde{C}_{\rm max}$. These features can be compared with our previous descriptions, thereby validating the generality of our analysis.}\label{fig:A3}
\end{figure*}

In \autoref{fig:A3} we plot the maximum wave speeds in more cases. Based on these and other exploration of the solution space, we find two transitions that identify weak and strong scattering limits. One transition is where $\widetilde{\sigma}=1/\widetilde{V}_m$.  If we define the CR mean free path $\lambda_{\rm mfp}\equiv c/\nu=(\sigma c)^{-1}$, then for wavelengths below this limit $1/k<\lambda_{\rm mfp}$ so that the CRs and the MHD gas fully decouple. Since we adopt maximum wave speed $\widetilde{V}_m/\sqrt{3}$ for the CR fluid, the fully decoupled limit is expected at $\widetilde{\sigma} \lesssim 1/\widetilde{V}_m$. The second transition appears roughly at $\widetilde{\sigma} = 1/\widetilde{C}_{\rm max}$, corresponding to the diffusion coefficient $\sigma^{-1}$ becoming 
smaller than $ C_{\rm max}\lambda$.  In this limit, all waves in the thermal gas become fully coupled, with negligible diffusion, to the CR fluid. 

Our analysis in the weak and strong scattering limits shows that solutions to the approximate dispersion relation have real $\widetilde{\omega}$ there, which is consistent with the small $|{\rm Im}(\widetilde{\omega})|$ found in the numerical solutions in those two regimes. When $\widetilde{\sigma}$ takes on an intermediate value, however, the assumption of a purely-real $\widetilde{\omega}$ clearly cannot satisfy the dispersion relation \autoref{eq:dispersion}. Thus, there must be a non-negligible ${\rm Im}(\widetilde{\omega})$. 

We observe that the numerical values of $|\mathrm{Re}(\omega)|$ are nearly identical to our asymptotic estimates, showing only a minute upward deviation. For numerical robustness, we multiply the asymptotic values by a safety factor of 2 in the HLLE solver, except for $\widetilde{V}_m/\sqrt{3}$, which would otherwise exceed the speed of light. Similarly, we adjust the two transition thresholds to $\widetilde{\sigma}=\sqrt{3}/\widetilde{V}_m$ and $\widetilde{\sigma}=1/(10\widetilde{C}_{\rm max})$, narrowing the intermediate regime and further overestimating the maximum wave speed used in the HLLE solver.

\section{Stability Criterion for Time Step}\label{append:timestep}

In this appendix, we go through the traditional von Neumann stability analysis to provide guidelines for the time step $\Delta t$. We make several simplifications in our analysis. First, since our goal is to study the numerical instability attributed to the sign-flipping feature of the scattering coefficient (\autoref{eq:stscatt}) at $P_{\rm cr}$ extrema under CRSI-driven scattering, we remove irrelevant terms including the Lorentz force terms and the $\Delta P_{\rm cr}$ terms in the equations. Furthermore, we set the CR energy source term zero in accordance with \autoref{eq:consEcr}. Finally, we omit the $-(\sigma_+ +\sigma_-)F_{\rm cr}$ term in the CR flux equation because it has no sign-flipping behavior and it is therefore not the cause of the problem. After all the reductions, and adopting the form for the scattering coefficient from \autoref{eq:stscatt} we are left with the two following equations (assuming static MHD gas and $\vec{b}=\hat{x}$):

\begin{equation}
    \frac{\partial \mathcal{E}_{\rm cr}}{\partial t}+\nabla\cdot\vec{F}_{\rm cr}=0
\end{equation}

\begin{equation}
    \frac{1}{V_m^2}\frac{\partial \vec{F}_{\rm cr}}{ \partial t}+\nabla\cdot\mathbf{P}_{\rm cr}=\frac{4}{3}(\sigma_+ - \sigma_-)\mathcal{E}_{\rm cr}v_A=\frac{4}{3}\widetilde{\sigma}_{0,\rm st}\frac{|\nabla \mathcal{E}_{\rm cr}|}{v_A \mathcal{E}_{\rm cr}}{\rm sgn}\left(-\nabla \mathcal{E}_{\rm cr}\cdot\vec{b}\right)v_A \mathcal{E}_{\rm cr}=-\frac{4}{3}\widetilde{\sigma}_{0,\rm st}\nabla \mathcal{E}_{\rm cr}
\end{equation}

We rewrite these equations by expanding advection terms $\nabla\cdot\vec{F}_{\rm cr}$ and $\nabla \cdot\mathbf{P}_{\rm cr}$ as we calculate them in our HLLE solver\footnote{Since the numerical instability is triggered near $P_{\rm cr}$ extrema where $\nabla P_{\rm cr}\rightarrow 0$, we directly plug $V^+=V_m/\sqrt{3}$ in this limit.} (\autoref{eq:HLLEflux}), and treating $\nabla \mathcal{E}_{\rm cr}$ on the RHS simply as central difference (the same way we calculate it in $\sigma_{\pm}^{L,R}$ in the code):
\begin{equation}
    \frac{E_{\rm cr,i}^{(n+1)}-E_{\rm cr,i}^{(n)}}{\Delta t}+\frac{F_{\rm cr,i+1}^{(n)}-F_{\rm cr,i}^{(n)}}{2\Delta x}-\frac{V_m}{2\sqrt{3}\Delta x}\left(E_{\rm cr,i+1}^{(n)}+E_{\rm cr,i-1}^{(n)}-2E_{\rm cr,i}^{(n)}\right)=0
\end{equation}

\begin{equation}
    \frac{F_{\rm cr,i}^{(n+1)}-F_{\rm cr,i}^{(n)}}{V_m \Delta t}+\frac{V_m}{6\Delta x}\left(E_{\rm cr,i+1}^{(n)}-E_{\rm cr,i-1}^{(n)}\right)-\frac{\left(F_{\rm cr,i+1}^{(n)}+F_{\rm cr,i-1}^{(n)}-2F_{\rm cr,i}^{(n)}\right)}{2\sqrt{3}\Delta x}=-\frac{4}{3}\widetilde{\sigma}_{0,\rm st}\frac{V_m}{2\Delta x}\left(E_{\rm cr,i+1}^{(n)}-E_{\rm cr,i-1}^{(n)}\right)
\end{equation}

Next we study one monochromatic mode such that $E_{\rm cr,i\pm 1}^{(n)}=e^{\pm ik\Delta x}E_{\rm cr,i}^{(n)}$ and $F_{\rm cr,i\pm 1}^{(n)}=e^{\pm ik\Delta x}F_{\rm cr,i}^{(n)}$. We suppose the amplification factor of this mode is $A_k$, meaning: $F_{\rm cr,i}^{(n+1)}/F_{\rm cr,i}^{(n)}=E_{\rm cr,i}^{(n+1)}/E_{\rm cr,i}^{(n)}=A_k$. Substituting back these relations we can eventually solve for the amplification factor

\begin{equation}
    A_k=1-\frac{2\sqrt{3}V_m \Delta t}{3\Delta x}{\rm sin}^2\frac{k\Delta x}{2}\pm i\sqrt{\frac{1+4\widetilde{\sigma}_{0,\rm st}}{3}}\frac{V_m \Delta t}{\Delta x}{\rm sin}k\Delta x
\end{equation}

Numerical stability requires $|A_k|^2<1$ for all $k$, which translates to

\begin{equation}
  -\frac{4\sqrt{3}}{3}\frac{V_m\Delta t}{\Delta x}{\rm sin}^2\frac{k\Delta x}{2}+\frac{4}{3}\frac{V_m^2 \Delta t^2}{\Delta x^2}{\rm sin}^4\frac{k\Delta x}{2}+\left(\frac{1+4\widetilde{\sigma}_{0,\rm st}}{3}\right)\frac{V_m^2 \Delta t^2}{\Delta x^2}{\rm sin}^2k\Delta x<0  
\end{equation}

The LHS of the inequality above equals 
\begin{equation}
    \frac{4}{3}\frac{V_m \Delta t}{\Delta x}{\rm sin}^2\frac{k\Delta x}{2}\left[\frac{V_m \Delta t}{\Delta x}+4\widetilde{\sigma}_{0,\rm st}\frac{V_m\Delta t}{\Delta x}{\rm cos}^2\frac{k\Delta x}{2}-\sqrt{3}\right]
\end{equation}

so the stability criterion becomes:
\begin{equation}
    \Delta t<\frac{\sqrt{3}\Delta x}{(1+4\widetilde{\sigma}_{0,\rm st})V_m}
\end{equation}

\begin{figure*}
    \centering
    \includegraphics[scale=0.41]{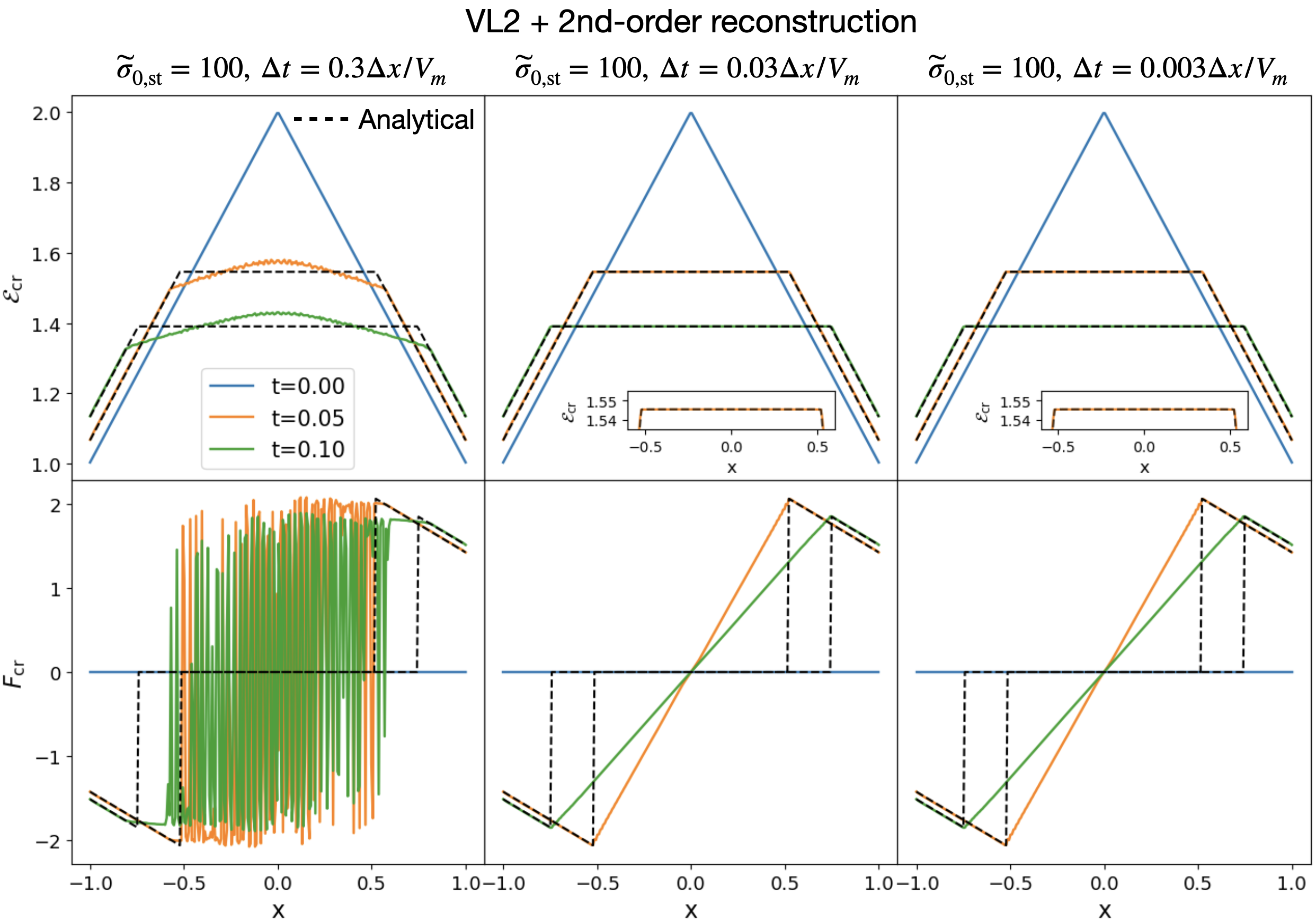}
    \caption{Performance of our numerical stability criterion \autoref{eq:stabcriterion} combined with the second-order scheme (VL2 integrator $+$ piecewise linear reconstruction). The rightmost column corresponds to time steps that exactly satisfy the stability condition. Compared to \autoref{fig:1Dstability}, which uses the RK1 integrator $+$ 1st-order reconstruction, we see the system is already stabilized in the middle column, demonstrating enhanced robustness.}
    \label{fig:1Dstability_vl2}
\end{figure*}

We emphasize that the derivation above corresponds to the numerical scheme combining first-order spatial reconstruction with an RK1 integrator, and the scattering coefficient should strictly adhere to \autoref{eq:stscatt}. For alternative numerical schemes or scattering mechanisms governed by distinct physics (e.g., Landau damping), analytical approaches can be developed in analogy, but we defer it to future studies. 

In practice, we adopt the VL2 integrator with second-order spatial reconstruction, as this configuration delivers enhanced numerical robustness and permits more relaxed time-step constraints. \autoref{fig:1Dstability_vl2} demonstrates tests analogous to those in \autoref{fig:1Dstability}, but using the VL2 integrator $+$ piecewise linear reconstruction. We see the numerical instabilities fade even when $\Delta t$ is up to an order of magnitude larger than the value required by our stability condition. We employ the derived stability criterion in all our streaming-dominated simulations, and also recommend it as a reference guideline and paradigm for similar studies.


\section{CR energy evolution in cylindrical coordinates with scattering due to CRPAI}\label{append:cylindricalCRPAI}

In this appendix, we provide a derivation for the $\mathcal{E}_{\rm cr}$ evolution in a cylindrical coordinate, where the CR scattering is subject to the CRPAI as described in \autoref{subsec:CRPAItest}. Specifically, we are trying to solve the two-moment equations:

\begin{equation}
    \frac{\partial \mathcal{E}_{\rm cr}}{\partial t}+\frac{1}{R}\frac{\partial (RF_{\rm cr})}{\partial R}=\left(\frac{4\widetilde{\sigma}_{0,\rm aniso}-1}{3}\right)\frac{1}{B}\left|\frac{dB}{dR}\right|\mathcal{E}_{\rm cr}v_A
\end{equation}

\begin{equation}
    \frac{1}{V_m^2}\frac{\partial F_{\rm cr}}{\partial t}+\frac{1}{3}\frac{\partial \mathcal{E}_{\rm cr}}{\partial R}=-\frac{\widetilde{\sigma}_{0,\rm aniso}}{v_A}\frac{1}{B}\left|\frac{dB}{dR}\right|F_{\rm cr}
\end{equation}
where $\widetilde{\sigma}_{0,\rm aniso}$ is a positive constant, $B(R)=B_0(R_0/R)$ and $v_A(R)=v_{A,0}(R_0/R)$ with $B_0$, $v_{A0}$ and $R_0$ being constants simply for normalizations.

Assuming the CR flux equation reaches a steady state, we have 
\begin{equation}
    F_{\rm cr}=-\frac{v_{A0}R_0}{3\widetilde{\sigma}_{0,\rm aniso}}\frac{\partial \mathcal{E}_{\rm cr}}{\partial R}
\end{equation}

plugging it back to the CR energy equation yields:

\begin{equation}
    \frac{\partial \mathcal{E}_{\rm cr}}{\partial t}-\frac{v_{A0}R_0}{3\widetilde{\sigma}_{0,\rm aniso}}\frac{\partial^2 \mathcal{E}_{\rm cr}}{\partial R^2}-\frac{v_{A0}R_0}{3\widetilde{\sigma}_{0,\rm aniso}R}\frac{\partial \mathcal{E}_{\rm cr}}{\partial R}=\left(\frac{4\widetilde{\sigma}_{0,\rm aniso}-1}{3}\right)\frac{v_{A0}R_0}{R^2}\mathcal{E}_{\rm cr}
\end{equation}

We do variable separation for $\mathcal{E}_{\rm cr}(t,R)=T(t)L(R)$, and after organizing terms we have:

\begin{equation}
    \frac{1}{T(t)}\frac{dT}{dt}=\frac{v_{A0}R_0}{3\widetilde{\sigma}_{0,\rm aniso}}\frac{1}{L(R)}\left[\frac{d^2 L}{dR^2}+\frac{1}{R}\frac{dL}{dR}+\left(\frac{4\widetilde{\sigma}_{0,\rm aniso}-1}{3R^2}\right)\widetilde{\sigma}_{0,\rm aniso}L\right]
\end{equation}

For temporal and spatial functions to equal, they must both equal to a constant, denoted as $-\lambda$. The temporal part has an obvious solution $T(t)=C_1e^{-\lambda t}$, while the spatial part satisfies the Bessel equation:

\begin{equation}
    R^2\frac{d^2L}{dR^2}+R\frac{dL}{dR}+\left(4\widetilde{\sigma}_{0,\rm aniso}^2-\widetilde{\sigma}_{0,\rm aniso}+\frac{3\widetilde{\sigma}_{0,\rm aniso}}{v_{A0}R_0}\lambda R^2\right)L=0
\end{equation}
whose general solution is:
\begin{equation}
    L(R)=C_2J_{\alpha}\left(\sqrt{\frac{3\lambda\widetilde{\sigma}_{0,\rm aniso}}{v_{A0}R_0}}R\right) + C_3Y_{\alpha}\left(\sqrt{\frac{3\lambda\widetilde{\sigma}_{0,\rm aniso}}{v_{A0}R_0}}R\right)
\end{equation}
where $J_{\alpha}$ and $Y_{\alpha}$ are respectively the first and second kind of Bessel functions. $\alpha\equiv\sqrt{(1-4\widetilde{\sigma}_{0,\rm aniso})\widetilde{\sigma}_{0,\rm aniso}}$ is the order, and $C_1$, $C_2$ and $C_3$ are integral constants.

In \autoref{subsec:CRPAItest}, for simplicity we choose $C_3=0$, $\widetilde{\sigma}_{0,\rm aniso}=0.25$, and $v_{A0}=R_0=\lambda=C_1=C_2=1$, so that one specific solution for $\mathcal{E}_{\rm cr}$ is $\mathcal{E}_{\rm cr}(t,R)=T(t)L(R)=J_0(\sqrt{3}R/2)e^{-t}$, which we send in as initial condition to test the performance our code.


\bibliography{sample631}{}
\bibliographystyle{aasjournal}



\end{document}